\documentclass[aps,prl,onecolumn,superscriptaddress,preprintnumbers,amsmath,amssymb,footinbib]{revtex4-2}
\pdfoutput=1 % for arxiv
\usepackage{xcolor}% coloured text
\usepackage[breaklinks=true,colorlinks,citecolor=blue,linkcolor=blue,urlcolor=blue]{hyperref}

\usepackage{chemformula} % Formula subscripts using \ch{}
\usepackage[T1]{fontenc} % Use modern font encodings
\usepackage{amsmath}
\usepackage{amsfonts}
\usepackage{color}
\usepackage{float}
\usepackage{tabularray}
\usepackage{tabularx}
\usepackage{multirow}
\usepackage{booktabs}
\usepackage{atbegshi}
\usepackage{upgreek}
\DeclareMathSymbol{\shortminus}{\mathbin}{AMSa}{"39}

\usepackage{xfp}

\newcommand\ExtendedDataFigures{%
	\xdef\presupfigures{\arabic{figure}}% save the current figure number
	\renewcommand{\figurename}{Extended Data Fig.}
	\renewcommand\thefigure{\fpeval{\arabic{figure}-\presupfigures}}
}

\newcommand\SupplementaryMaterials{%
	\xdef\presupfigures{\arabic{figure}}% save the current figure number
		\renewcommand{\figurename}{Supplementary Fig.}
}

\begin{document}
	
	\title[  ]{Ultrafast Umklapp-assisted electron-phonon cooling in magic-angle twisted bilayer graphene}
	
	\author{Jake Dudley \surname{Mehew}} 
	\affiliation{Catalan Institute of Nanoscience and Nanotechnology (ICN2), BIST and CSIC, Campus UAB, 08193 Bellaterra (Barcelona), Spain}

	\author{Rafael \surname{Luque Merino}}
	\affiliation{ICFO - Institut de Ciencies Fotoniques, The Barcelona Institute of Science and Technology (BIST), Castelldefels 08860, Spain}
	\affiliation{Fakult\"{a}t f\"{u}r Physik, Ludwig-Maximilians-Universit\"{a}t, Schellingstrasse 4, M\"{u}nchen 80799, Germany}
	\affiliation{Munich Center for Quantum Science and Technology (MCQST), M\"{u}nchen, Germany}
	
	\author{Hiroaki \surname{Ishizuka}}
	\affiliation{Department of Physics, Tokyo Institute of Technology, Tokyo, Japan}
		
	\author{Alexander \surname{Block}}
	\affiliation{Catalan Institute of Nanoscience and Nanotechnology (ICN2), BIST and CSIC, Campus UAB, 08193 Bellaterra (Barcelona), Spain}

	\author{Jaime \surname{D\'{i}ez M\'{e}rida}}
	\affiliation{ICFO - Institut de Ciencies Fotoniques, The Barcelona Institute of Science and Technology (BIST), Castelldefels 08860, Spain}
	\affiliation{Fakult\"{a}t f\"{u}r Physik, Ludwig-Maximilians-Universit\"{a}t, Schellingstrasse 4, M\"{u}nchen 80799, Germany}
	\affiliation{Munich Center for Quantum Science and Technology (MCQST), M\"{u}nchen, Germany}

	\author{Andr\'{e}s \surname{D\'{i}ez Carl\'{o}n}}
	\affiliation{ICFO - Institut de Ciencies Fotoniques, The Barcelona Institute of Science and Technology (BIST), Castelldefels 08860, Spain}
	\affiliation{Fakult\"{a}t f\"{u}r Physik, Ludwig-Maximilians-Universit\"{a}t, Schellingstrasse 4, M\"{u}nchen 80799, Germany}
	\affiliation{Munich Center for Quantum Science and Technology (MCQST), M\"{u}nchen, Germany}
	
	\author{Kenji \surname{Watanabe}}
	\affiliation{Research Center for Functional Materials, National Institute for Material Sciences, Tsukuba, Japan}
	
	\author{Takashi \surname{Taniguchi}}
	\affiliation{International Center for Materials Nanoarchitectonics, National Institute for Material Sciences, Tsukuba, Japan}
	
	\author{Leonid S. \surname{Levitov}}
	\affiliation{Department of Physics, Massachusetts Institute of Technology, Cambridge, 02139 MA, USA}
	
	\author{Dmitri K. \surname{Efetov}}
	\affiliation{Fakult\"{a}t f\"{u}r Physik, Ludwig-Maximilians-Universit\"{a}t, Schellingstrasse 4, M\"{u}nchen 80799, Germany}
	\affiliation{Munich Center for Quantum Science and Technology (MCQST), M\"{u}nchen, Germany}
	
	\author{Klaas-Jan \surname{Tielrooij}}\email[Correspondence to: ]{klaas.tielrooij@icn2.cat} 
	\affiliation{Catalan Institute of Nanoscience and Nanotechnology (ICN2), BIST and CSIC, Campus UAB, 08193 Bellaterra (Barcelona), Spain}
	\affiliation{Department of Applied Physics, TU Eindhoven, Den Dolech 2, Eindhoven, 5612 AZ, The Netherlands.}
	
	\begin{abstract}
		\vspace{0.5cm}
	Carrier relaxation measurements in moir\'{e} materials offer a unique probe of the microscopic interactions, in particular the ones that are not easily measured by transport. Umklapp scattering between phonons is a ubiquitous momentum-nonconserving process that governs the thermal conductivity of semiconductors and insulators. In contrast, Umklapp scattering between electrons and phonons has not been demonstrated experimentally. Here, we study the cooling of hot electrons in moir\'e graphene using time- and frequency-resolved photovoltage measurements as a direct probe of its complex energy pathways including electron-phonon coupling. We report on a dramatic speedup in hot carrier cooling of twisted bilayer graphene near the magic angle: the cooling time is a few picoseconds from room temperature down to 5 K, whereas in pristine graphene coupling to acoustic phonons takes nanoseconds. Our analysis indicates that this ultrafast cooling is a combined effect of the formation of a superlattice with low-energy moir\'{e} phonons, spatially compressed electronic Wannier orbitals, and a reduced superlattice Brillouin zone, enabling Umklapp scattering that overcomes electron-phonon momentum mismatch. These results demonstrate a way to engineer electron-phonon coupling in twistronic systems, an approach that could contribute to the fundamental understanding of their transport properties and enable applications in thermal management and ultrafast photodetection.
	\end{abstract}
	
	\maketitle
	\section{Introduction}\label{sec1}
	Moir\'{e} superlattices provide a novel material platform in which twist angle controls the effective lattice constant. As the twist angle decreases, the larger moir\'{e} unit cell corresponds to a smaller electron momentum. This tunes the relative strength of the kinetic energy of electrons and the interaction energy between them. In magic-angle twisted bilayer graphene (MATBG), these interactions result in a rich phase diagram that includes superconductors, \cite{Cao2018,Lu2019,Yankowitz2019,Stepanov2020} correlated insulators \cite{Cao2018b,Wong2020} and orbital magnets. \cite{Lu2019,Sharpe2019} In transition metal dichalcogenides (TMDs), correlated insulating \cite{Regan2020,Tang2020} and ferromagnetic states \cite{Wang2022} are observed over a broad range of angles, with moir\'{e} excitons \cite{Seyler2019,Alexeev2019} providing a test-bed for exploring Hubbard model physics. \cite{Tang2020} In addition, the moir\'{e} potential modifies the phonon spectra for small twist angles. \cite{Lin2018} This results in phonon renormalization in MoS$_2$ homobilayers \cite{Quan2021} and the emergence of phonon minibands in twisted bilayer graphene. \cite{Koshino2019} Theoretical studies predict that the moir\'{e} potential strongly affects electron-phonon coupling, \cite{Wu2018,Choi2018,Koshino2020} which has important implications for electrical transport, excited-state relaxation dynamics, and beyond. 
	
	Excited-state relaxation measurements are particularly well-suited probes to quantitatively assess electron-phonon coupling. The relaxation dynamics in graphene after excitation involve thermalization of high-energy carriers through carrier-carrier scattering within tens of femtoseconds, \cite{Tielrooij2015} creating a hot carrier distribution that subsequently cools via phonons. Inelastic electron-phonon scattering allows electrons to gain (lose) energy by the absorption (emission) of a phonon. In graphene, cooling typically occurs via the emission of optical and acoustic graphene phonons, and near-field coupling to substrate phonons. \cite{Bistritzer2009,Song2012,Graham2013,Kong2018,Tielrooij2018,Massicotte2021,Pogna2021a,Kim2021} Importantly, in all cases, cooling becomes increasingly slow for lower lattice temperatures, with predicted electron-phonon cooling times in the nanosecond regime for the case of pristine graphene at cryogenic temperatures. \cite{Bistritzer2009} 
	
	Experimental studies of the relaxation dynamics of twisted bilayer graphene have so far been limited to large twist angles ($\theta>5^\circ$). In these systems, a dark exciton state emerges between van Hove singularities, leading to slower dynamics. \cite{Patel2015,Patel2019} At such relatively large angles, the moir\'{e} potential has limited influence on electron-phonon coupling. \cite{Koshino2019,Koshino2020} Recent Raman spectroscopy measurements suggest an enhanced electron-phonon coupling strength for small twist angles around the magic angle ($\theta\approx1.1^\circ$). \cite{Gadelha2021} However, direct experimental measurements of moir\'{e}-enhanced electron-phonon coupling and its implications for cooling dynamics, are lacking, nor is there any clear understanding of the origin of the enhanced coupling. 
	
\begin{figure}[ht!]%
		\centering
		\includegraphics[width=\textwidth]{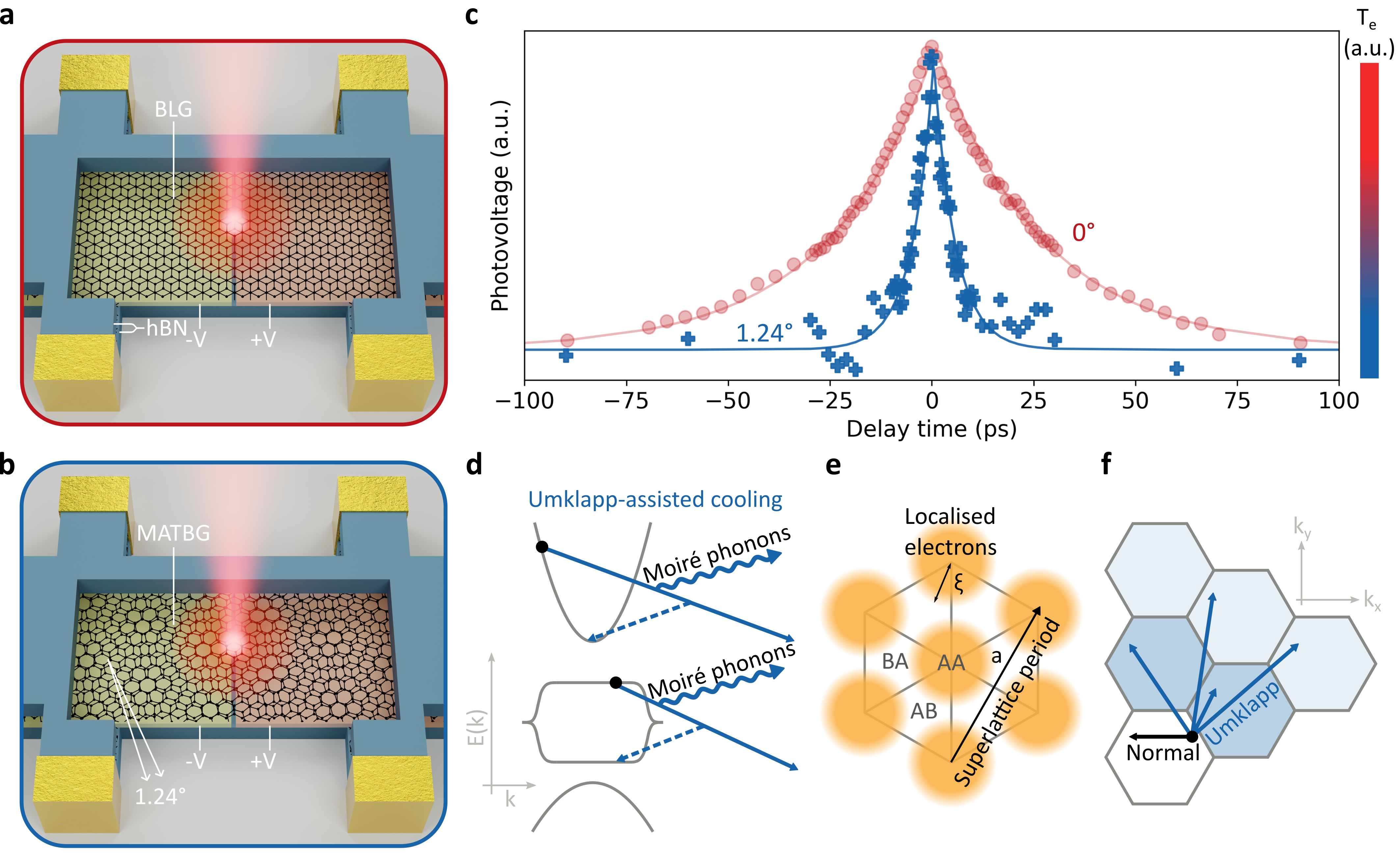}
		\caption{\textbf{Excited carrier relaxation in MATBG.} \textbf{a-b}, Illustration of the hBN-encapsulated BLG device with $0^\circ$ twist angle \textbf{(a)} and the hBN-encapsulated MATBG device with twist angle $1.24^\circ$ \textbf{(b)}, each equipped with split gates. By applying voltages of opposite sign ($\pm V$) to the split gates, we create a pn-junction (the interface between yellow and orange regions). Illuminating the junction generates a photovoltage via the photothermoelectric effect, which is proportional to the electron temperature ($T_e$). We obtain the temperature dynamics either by using two ultrashort laser pulses separated in time by a variable temporal delay, \cite{Urich2011,Sun2012,Tielrooij2015} or by using two spectrally narrow laser beams with variable frequency detuning. \cite{Jadidi2016,Aamir2021} \textbf{c}, Photovoltage as a function of time delay for a lattice temperature of 25 K. The decay, which represents the cooling dynamics, is much faster in MATBG (blue pluses) than BLG (red circles). 
			\textbf{d}, Schematic of the MATBG band structure. Umklapp scattering processes (solid arrow) allow for efficient electron (black circle) relaxation via coupling to moir\'{e} phonons (wiggly lines). These Umklapp processes can occur in both the flat and the dispersive bands. The dashed arrows represent the equivalent final state in the first Brillouin zone. 
			\textbf{e}, Schematic of the compressed Wannier orbitals of radius $\xi$. Electrons are localised to AA sites in the reconstructed superlattice. 
			\textbf{f}, Umklapp scattering processes (blue arrows) couple electrons in the first Brillouin zone (white hexagon) to large-momentum phonons in higher-order Brillouin zones (blue hexagons).} \label{fig1}
	\end{figure}

	In this paper, we report the observation of ultrafast cooling in magic-angle twisted bilayer graphene (MATBG) through Umklapp-assisted electron-phonon scattering. We directly probe the electron-phonon interaction by measuring carrier cooling dynamics using two well-established optoelectronic techniques – time-resolved photovoltage microscopy \cite{Urich2011,Sun2012,Tielrooij2015} and continuous-wave photomixing \cite{Jadidi2016,Aamir2021}. We make a direct comparison between a non-twisted Bernal bilayer graphene sample (BLG, $\theta=0^\circ$, see Fig.\ \ref{fig1}a) and a near-magic-angle twisted bilayer graphene sample (MATBG, $\theta=1.24^\circ$, see Fig. \ref{fig1}b). At low temperature, the cooling dynamics are much faster in MATBG than in non-twisted bilayer graphene, see Fig. \ref{fig1}c. This unexpected result highlights the crucial role of the moir\'{e} pattern and suggests the emergence of an enhanced electron-phonon interaction in small twist angle systems. We explain the observed relaxation dynamics using a theoretical model based on Umklapp-assisted electron-phonon scattering, which can occur in both the dispersive and flat bands of MATBG, see Fig. \ref{fig1}d. The Umklapp processes are enabled by the presence of compressed electronic Wannier orbitals (see Fig.\ref{fig1}e), and the superlattice with reduced Brillouin zone (see Fig. \ref{fig1}f).
	
	\begin{figure}[ht]%
		\centering
		\includegraphics[width=\textwidth]{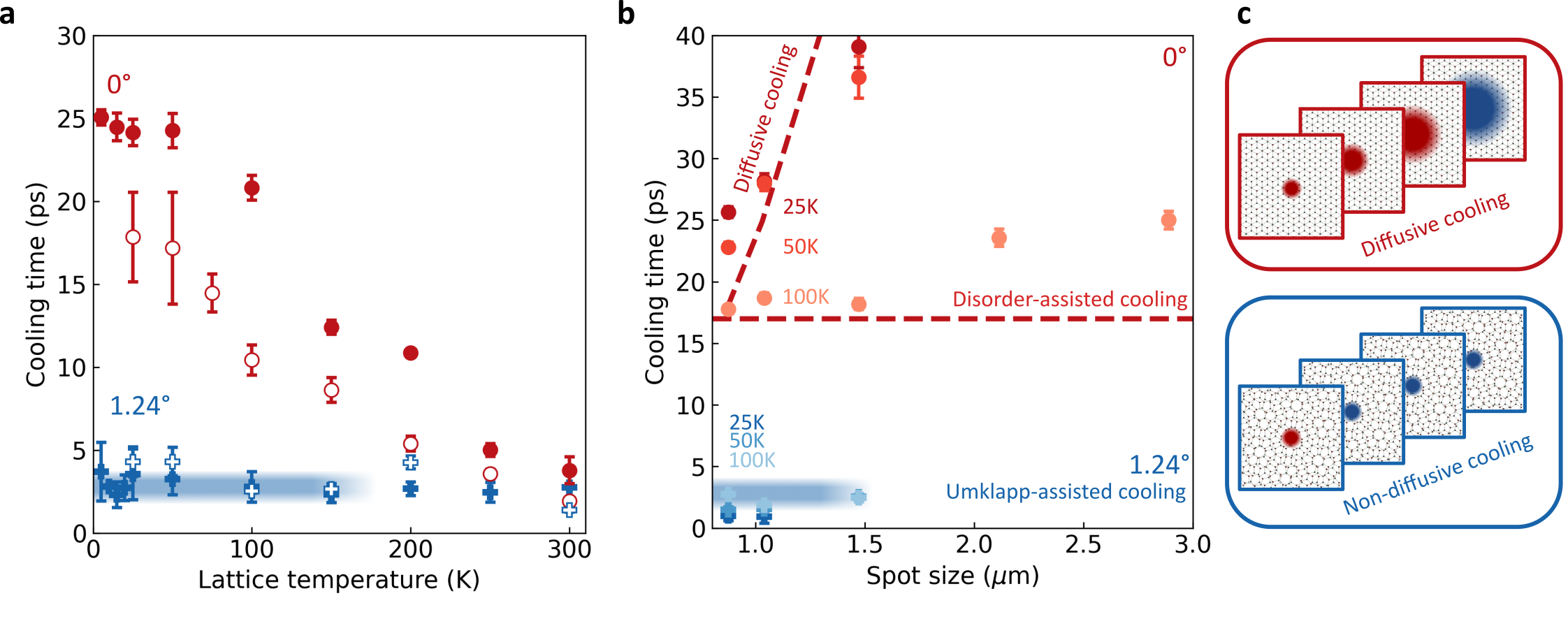}
		\caption{\textbf{Relaxation mechanisms in MATBG and BLG.} \textbf{a}, Cooling time as a function of lattice temperature. In MATBG ($1.24^\circ$, blue pluses), the cooling time is constant between 5 K and 300 K (3 ps, blue line). For BLG ($0^\circ$, red circles), it is greater at lower temperatures. \textbf{b}, Laser spot size dependence of the cooling time. The strong dependence in BLG at 25 K and 50 K is a signature of diffusive cooling. This effect is weaker at 100 K, where disorder-assisted cooling becomes significant. The effect is absent in MATBG for these spot sizes. The filled (open) shapes are measured using the TrPV (CW-PM) technique. Error bars represent the statistical spread across different gate voltages. The thick blue line in \textbf{a} and \textbf{b} represents the cooling time obtained from the low temperature model of Umklapp-assisted cooling (see Main text). \textbf{c} Schematics of diffusive cooling for BLG (upper) and its absence for MATBG (lower).}\label{fig2}
	\end{figure}
	
	\section{Relaxation dynamics}
	We study relaxation dynamics in hBN-encapsulated MATBG and BLG Hall bar devices as shown in Fig. \ref{fig1}a-b (see Methods for details on the device fabrication and characterization). These devices enable both electrical and optoelectronic measurements, as they are equipped with a split gate that we employ to create a photoactive pn-junction region. The resistance map as a function of the gate voltage applied to each of the two sides of the split gate, shown in Extended Data Figure \ref{edf1}, displays clear peaks at the usual Dirac points with vanishing carrier density. The MATBG device exhibits additional peaks at integer fillings of the superlattice unit cell. By illuminating the pn-junction with light, a photovoltage is generated via the photothermoelectric effect. This effect has a characteristic six-fold symmetry in dual-gate photovoltage maps, as shown in Extended Data Fig. \ref{edf2} for both devices. This indicates that the measured photovoltage is a direct probe of the electron temperature. \cite{Gabor2011}
	
	We study hot electron relaxation using ultrafast time-resolved photovoltage microscopy (TrPV) as implemented in Refs. \cite{Tielrooij2015,Tielrooij2018} and continuous-wave heterodyne photomixing (CW-PM) as implemented in Ref. \cite{Aamir2021} In the former, ultrashort laser pulses are incident upon the pn-junction whereas for the latter two continuous wave lasers are used. In both cases we probe the generated photovoltage. These two techniques allow us to obtain directly the carrier cooling dynamics -- in the time domain by varying the time delay between two ultrashort laser pulses, and in the frequency domain by varying the spectral detuning of two spectrally narrow laser beams. Both techniques independently show that charge carriers cool much faster in MATBG than in BLG at low temperature, see Fig. \ref{fig2}a (and Extended Data Figs. \ref{edf3}-\ref{edf6}). In BLG the cooling time increases from 3 ps to 25 ps as the temperature decreases from 300 K to 5 K, which is expected as it takes longer for hot carriers to couple to phonons at lower temperature due to the reduced phonon occupation. \cite{Bistritzer2009,Song2012} Surprisingly, for MATBG the cooling time remains short, around 3 ps, across a broad temperature range (5-300 K). This suggests the involvement of low-energy phonons that still have occupation at such low temperature, which are likely phonons related to the superlattice. Indeed, the moir\'{e} potential breaks the original linear phonon dispersion into minibands with enhanced density of states. \cite{Koshino2019} The energy of the lowest band is below 1 meV corresponding to temperatures below 10 K. 
	
	In order to understand the origin of the observed cooling dynamics, we first consider the case of relaxation through energy transfer to phonons in non-twisted BLG. Coupling to optical phonons is highly inefficient at low temperature due to the large optical phonon energy, which is $>160$ meV, corresponding to $T>2000$ K. \cite{Pogna2021a} Coupling between electrons and acoustic phonons is normally also inefficient due to the reduced phase space available for scattering, and would give cooling times well above a nanosecond below 25 K. \cite{Bistritzer2009} The presence of defects can help overcome the electron-phonon momentum mismatch through disorder-assisted cooling, which speeds up this acoustic phonon cooling process. \cite{Song2012,Graham2013,Kong2018} However, even with this mechanism, we expect cooling times between $10^{-10}$ s and $10^{-8}$ s for the lowest temperatures, depending on the electron mean free path (see Methods). We therefore consider diffusive cooling, where electronic heat diffuses out of the initially excited hot spot, thus leading to a lower average electron temperature. \cite{Massicotte2021} In this diffusive cooling mechanism, the cooling time will thus depend on laser spot size. Indeed, for non-twisted BLG, we observe an increase in cooling time for larger spot sizes, which is largest for the lowest temperatures (25 K, 50 K), see Fig. \ref{fig2}b-c. At 100 K, the cooling length is shorter and therefore diffusive cooling has a smaller contribution. We thus understand the cooling dynamics for non-twisted BLG from a combination of disorder-assisted and diffusive cooling. Indeed, our calculations of the cooling time based on these two mechanisms are close to the experimentally observed ones (see Methods for details on the calculations). Importantly, for MATBG we observe no dependence of the cooling time on spot size (see Fig. \ref{fig2}b-c), which suggests that diffusive cooling does not play a role for this system. 
	
\begin{figure}[ht]%
		\centering
		\includegraphics[width=0.5\textwidth]{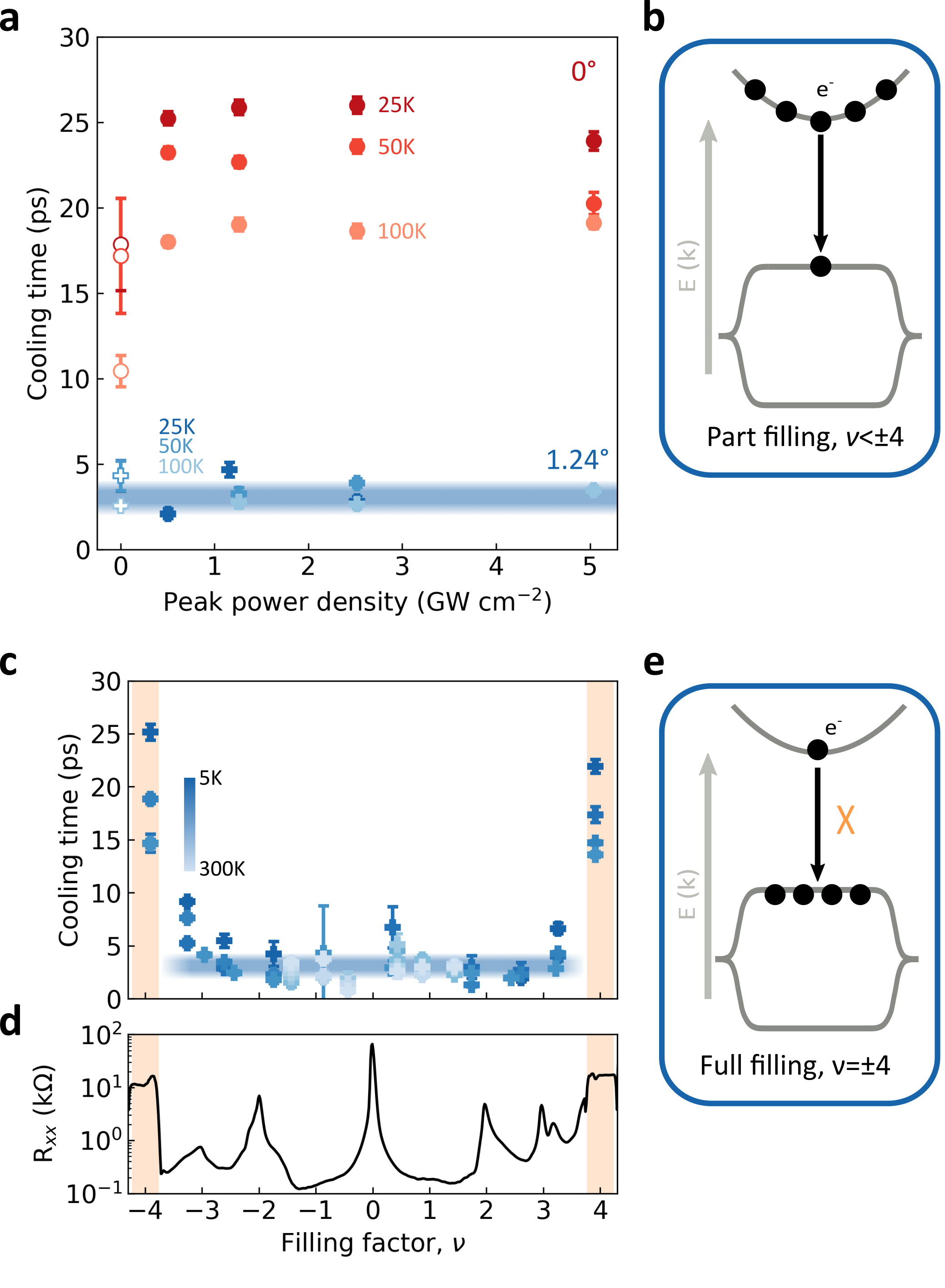}
		\caption{\textbf{Origin of enhanced cooling in MATBG.} \textbf{a}, Dependence of cooling time on peak power density for BLG (red circles) and MATBG (blue pluses). The filled (open) shapes are measured using the TrPV (CW-PM) technique. The error bars signify the one sigma confidence interval from the fitting algorithm. \textbf{b, e} Schematics of cooling power in MATBG for part filling (\textbf{b}) and full filling (\textbf{e}) of the flat bands. For part filling, the interband transition is not rate-limiting as evidenced by the absence of a power dependence in \textbf{a}. At full filling, cooling times are longer due to the interband bottleneck effect illustrated in panel \textbf(e). \textbf{c-d}, Gate dependence of cooling time, \textbf{c}, and four terminal resistance acquired at T = 35 mK ($R_{xx}$), \textbf{d}. Orange shaded region highlights full filling of the moir\'{e} unit cell, where $R_{xx}$ and cooling time increase. The thick blue line in \textbf{a} and \textbf{c} represents the cooling time obtained from the low temperature model of Umklapp-assisted cooling (see Main text)}.	\label{fig3}
	\end{figure}

	We next study the effect of changing the laser power and therefore initial electron temperature, see Fig. \ref{fig3}a. This corresponds to increasing the population of the dispersive band (Fig. \ref{fig3}b). The peak power density is roughly five orders of magnitude larger for our pulsed laser experiment (TrPV) than our continuous wave experiment (CW-PM). For the non-twisted BLG device, we observe somewhat slower cooling at higher initial electron temperature, which has also been observed for high-quality monolayer graphene samples and was ascribed to a bottleneck involving optical and acoustic phonons. \cite{Pogna2021a} Interestingly, the role of electron temperature is minor for the relaxation dynamics of MATBG, suggesting that there are no electron-phonon or phonon-phonon bottlenecks. This result indicates that direct optical phonon emission does not play a role in MATBG. The much faster cooling for MATBG compared to BLG at low temperatures thus suggests that a completely different mechanism is responsible for cooling, outcompeting all currently known cooling mechanisms in non-twisted graphene.
	
	We therefore explore the effect of the superlattice on electron cooling by examining the cooling time in MATBG as a function of filling factor ($\nu$) which represents the electronic occupation of the superlattice unit cell. For most filling factors ($\lvert\nu\rvert<4$), we observe a nearly constant cooling time of 3 ps across a wide temperature range (5-300 K). However, at $\nu=\pm4$ the cooling time increases dramatically. Low-temperature transport measurements on the same device reveal an increase in resistance at the same voltages, see Fig. \ref{fig3}d, which confirms the full filling of the superlattice unit cell. In Extended Data Figure \ref{edf7}, we show that the cooling time increases strongly upon increasing laser power at full filling for a second MATBG device ($\theta=1.08^\circ$). The strong dependence of cooling upon the flat band  filling - with the cooling rates high at partial filling and lower at full filling - indicates that the moir\'{e} pattern and its low-energy phonons are crucial for explaining the ultrafast cooling dynamics observed in MATBG. 
	
	\section{Origin of enhanced cooling}
	
	\begin{figure}[ht]%
		\centering
		\includegraphics[width=0.5\textwidth]{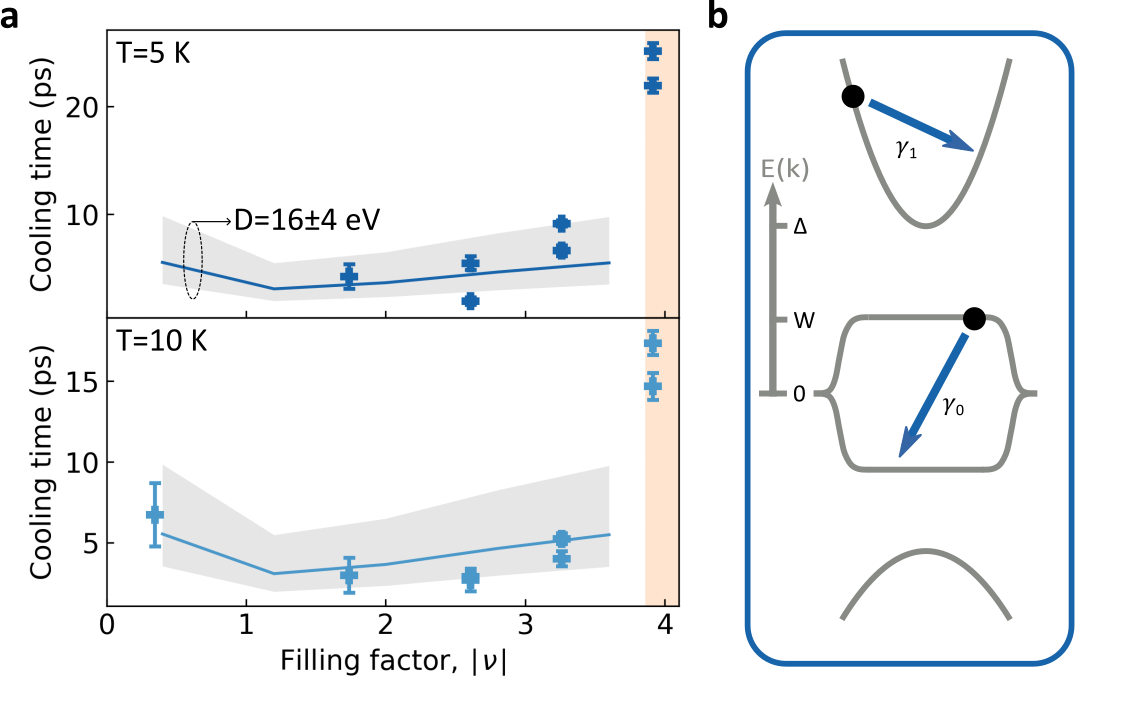}
		\caption{\textbf{Quantitative comparison with Umklapp-assisted cooling.} \textbf{a}, Comparison between calculated (solid line) and experimental (symbols) cooling times for MATBG at 5 K and 10 K (upper and lower panels). The grey shaded region allows for uncertainty in the value of the deformation potential ($D=16\pm4$ eV). 
			\textbf{b}, Schematic of the model used for the calculations with two dispersive and two flat bands separated by an energy gap ($\Delta - W$). $\gamma_1$ and $\gamma_0$ represent
			intra-dispersive-band and intra-flat-band scattering processes, respectively. The low temperature calculations shown in (\textbf{a}) consider only $\gamma_0$.}\label{fig4}
	\end{figure}

To gain insight into the different mechanisms that govern electron-lattice cooling pathways, we consider in detail the microscopic electron-phonon scattering processes in the MATBG system. To this end, we consider a four-band model consisting of two nearly-flat and two dispersive bands (Fig.~\ref{fig1}d). There are two main types of electron-phonon scattering in this model, interband and intraband. The intraband processes for the intra-dispersive band and intra-flat-band transitions are different and must be evaluated separately. At temperatures higher than the bandgap, which corresponds to the highest temperatures in our measurement, the electrons are thermally excited to the dispersive bands allowing both dispersive and flat bands to contribute to cooling. To the contrary, when the electron temperature is low, all carriers reside in the flat band. Therefore, we consider two regimes: i) the high temperature regime ($T\sim 150-300$ K), where the dispersive bands contribute to the cooling process, and ii) the low temperature regime ($T\sim 10$ K), wherein cooling is dominated by intra-flat-band processes. In both cases, we consider both the Umklapp and normal scattering contributions, finding that at the temperatures of interest ($T>10$ K) Umklapp scattering consistently wins over normal scattering.

For the first regime (high temperatures) we consider a four-band model consisting of two flat bands of bandwidth $W$ and two dispersive bands with the eigenstate energies $\varepsilon>\Delta$ and $\varepsilon<-\Delta$ ($\Delta> W$), see Fig. \ref{fig4}b. The dispersive bands are separated from the flat bands by a gap $\Delta-W$ (see Methods for details). A direct analysis based on Boltzmann theory yields cooling rates dominated by the intra-band processes in the dispersive bands, whereas the interband processes have a minor contribution. Accounting for the Umklapp processes, we estimate the cooling rates as $\tau^{-1}=\frac{6\rho_1}{\pi T_{el}}\sum_m(\|g^{1,1}_m\|^2+\|g^{-1,-1}_m\|^2)\omega_m^2$, where $\rho_1$ is the density of states of the dispersive particle and hole bands labeled by $n=\pm 1$, $T_{el}$ is the electron temperature, $g^{n,n}_m$ is the electron-phonon coupling constant in the $n^\text{th}$ band and $\omega_m$  is the phonon energy in the $m^\text{th}$ phonon band. Direct calculation gives cooling rates that are independent of the lattice temperature $T_{ph}$, in agreement with the observed dynamics, see Fig. \ref{fig2}a.

For the regime of low temperatures, we describe the system using a model of a flat band with electron and hole subbands (see Methods for a detailed description of the model). For a quantitative comparison with the experimental results shown in Fig.~\ref{fig4}a, we calculate the cooling power $J$ accounting for the Umklapp processes assuming the Wannier function radius $\xi=a/6$ where $a$ is the lattice parameter for the moir\'e structure~\cite{Ishizuka2021}, see Fig. \ref{fig1}c. The cooling rate $\tau^{-1}$ is estimated from the calculated cooling power and specific heat using  $\tau^{-1}=J/C(T_{el}-T_{ph})$;
here we calculate the specific heat $C$ using the fluctuation formula, Eq.~\ref{eq:C} in the Methods section. In that temperature values are not constrained by the flat-band width and can be as large as the bandgap. The filling dependence of the cooling rate is shown in Fig. \ref{fig4}a. The calculated Umklapp-assisted cooling times as a function of the filling factor are seen to be in agreement with the experimental results. For the calculated cooling times, we used a deformation potential of 16 eV. This is close to the values reported for single-layer graphene (10-30 eV). \cite{Efetov2010,Chen2008,Graham2013,Dean2010} We therefore conclude that the Umklapp-assisted carrier cooling model reproduces the main experimental findings. 

We note that, here, we did not take account of the disorder-assisted cooling processes.~\cite{Song2012} In pristine graphene, the bottleneck due to limited phase space due to the small Fermi surface is relieved by disorder scattering. The situation in MATBG differs from that in pristine graphene in two ways. First, as the superlattice provides additional momentum recoil, MATBG does not require defects and/or disorder for electron-lattice cooling. Second, the formation of highly localized Wannier orbitals at AA sites in the moir\'{e} pattern modulates the electron-phonon interaction. These  effects produce strong coupling of the electrons to moir\'e phonons even in the absence of disorder.~\cite{Ishizuka2021} 

\section{Outlook}
Importantly, the cooling measurement is predominantly sensitive to the electron-phonon interactions, and is less sensitive to the electron-electron interactions. This presents a unique window of opportunity for probing underlying physics, and an advantage compared to other measurements types that do not easily separate these two interactions. The finding that electron-phonon Umklapp scattering dominates ultrafast electron-phonon cooling is likely to have important implications for MATBG physics. 
Electron-phonon scattering plays an important role in charge transport, limiting the carrier mobility at high temperatures. This interaction also mediates the pairing interaction in Bardeen-Cooper-Schrieffer superconductors. Understanding the electron-phonon coupling could give important insights into the origin of superconductivity in MATBG. \cite{Wu2018,Peltonen2018}
For metals, electron-electron Umklapp scattering gives rise to finite electrical resistance at low temperatures. In graphene/hBN superlattices and MATBG, this effect dominates transport at temperatures up to 10 K or higher, leading to excess resistivity and degradation of charge carrier mobility. \cite{Wallbank2019,Jaoui2022,Ishizuka2022}
In MATBG, electron-phonon Umklapp scattering could explain some of the open questions from electrical transport measurements, such as the strange metal phase or the role of phonons in superconductivity.~\cite{Wu2018,Peltonen2018} 
Finally, the ultrafast Umklapp-assisted electron-phonon cooling, enhanced density of states, and rich phase diagram are appealing for single-photon detection in the highly sought after mid-IR wavelength range. \cite{DiBattista2022,Deng2020} 

\section{Methods}\label{sec11}
\textbf{Device fabrication}
The MATBG devices were fabricated using a cut and stack technique. All flakes were first exfoliated on a Si/SiO$_2$ (285 nm) substrate and later picked up using a polycarbonate (PC)/polydimethylsiloxane (PDMS) stamp. All the layers were picked up at a temperature of $\sim100^\circ$C. We used an AFM tip to cut the graphene in order to avoid strain during the pick-up process. The PC/PDMS stamp picks up first the top graphite layer, the top hBN and the first graphene layer. Before picking up the second graphene layer, we rotate the stage by an angle of $1.1-1.2 ^\circ$. Finally, the stamp picks up the bottom hBN and bottom graphite gates. We drop the finalized stack on a Si/SiO$_2$ substrate by melting the PC at $180^\circ$C, see Supplementary Figure S1a. The resulting stack is etched into a Hall bar using a CHF$_3$/O$_2$ plasma and a 1D contact is formed by evaporating Cr (5 nm)/Au (50 nm), see Supplementary Figure S1b. We etch a narrow channel of $\sim150$ nm in the top gate using an O$_2$ plasma. Before etching the top gate, the device was characterized at $T=35\textrm{ mK}$ to identify the pair of contacts closest to the magic angle ($\theta\sim1.1^\circ$). The junction was made in between this pair of contacts.

\textbf{Twist angle extraction}
The twist angle $\theta$ is extracted from the superlattice carrier density of the full band $n_s$ by applying the relation $n_s=8\theta^2/\sqrt{3}a^2$, where $a=0.246\textrm{ nm}$ is the graphene lattice constant. First, we calibrate the gate induced carrier density using the Hall effect data at $\pm1\textrm{ T}$. In the carrier density region close to charge neutrality, the Hall carrier density $n_H=-B/eR_{xy}$ should closely follow the gate induced carrier density $n_H=n$, see Supplementary Figure S2. By plotting $n_H$ vs $V_g$ and fitting this slope around charge neutrality we can obtain the capacitance of the device and therefore extract the real carrier density n. Then we extract the carrier density corresponding to a fully filled superlattice unit cell, in this case we find it to be $n_s=(3.58\pm 0.10)\times 10^{12} \textrm{ cm}^{-2}$. Finally using the above relation we extract a twist angle $\theta=1.24^\circ\pm0.02^\circ$. In Supplementary Note 1, we verify that there is minimal twist angle disorder in the junction region.

\textbf{Transport Measurements}
Low-temperature transport measurements were carried out in a dilution refrigerator (Bluefors SD250) with a base temperature of 20 mK. Standard low-frequency lock-in techniques (Stanford Research SR860 amplifiers) were used to measure $R_{xx}$ with an excitation current of 10 nA at a frequency of 13.11 Hz. 

\textbf{Optoelectronic measurements}
In time-resolved photovoltage (TrPV) experiments, we vary the delay time ($dt$) between the arrival of two ultrafast pulses. \cite{Urich2011} \cite{Sun2012} \cite{Tielrooij2015} Due to the non-linear relationship between carrier temperature and optical heating, we observe a dip in the photovoltage when the two pulses arrive at the same time ($dt=0$), see Fig. \ref{fig1}c and Extended Data Figs. \ref{edf3} and \ref{edf4}. At longer delay times, the signal recovers to its maximal value. We obtain the cooling time by describing the observed dynamics with an exponential function. 
For heterodyne photomixing (CW-PM) experiments, the wavelength detuning between the two continuous wave lasers creates an optical beating. \cite{Jadidi2016,Aamir2021} The photovoltage oscillates at the beating frequency. Due to the competition between beat frequency ($\Omega$) and the characteristic cooling time ($\tau_e$), we observe a peak for $\Omega=0$ whereas the oscillations are damped when $\Omega^{-1}\ll\tau_e$, see Extended Data Figs. \ref{edf5} and \ref{edf6}. 
%By describing the frequency response with a Lorentzian function of width $\Gamma$, we obtain the cooling time using the expression: $\Gamma=1/\pi\tau$. \cite{Jadidi2016}
The frequency response takes the form of a Lorentzian function of width $\Gamma$, from which we extract the cooling time as: $\Gamma=1/\pi\tau_e$. \cite{Jadidi2016}

\textbf{Estimating cooling times in untwisted graphene}
The hot electron cooling time for energy transfer to acoustic phonons in monolayer graphene is given by $\tau_{AP}\approx848/(D^2 T_L^2)$ [$\mu$s], \cite{Bistritzer2009} where $D$ is the deformation potential in eV. This expression is valid in the neutral limit ($T_F<T_e$) and close to equilibrium ($T_e\gtrsim T_L$). $T_{e/L/F}$ is the electron/lattice/Fermi temperature. \cite{Bistritzer2009} Taking $D=20 \textrm{ eV}$, we calculate a cooling time of $\tau_{AP}=3.4 \textrm{ ns}$ for $T_L=25 \textrm{ K}$. 

In disorder-assisted or supercollision cooling, \cite{Song2012,Graham2013,Kong2018} the dependence on lattice temperature is given by:
\begin{align}
&\tau_{SC} = \frac{\alpha}{3AT_L} \textrm{, with} \nonumber \\
&\alpha = \frac{2\pi E_F k_B^2}{3\hbar^2 v_F^2}  \textrm{ and } A=9.62 \frac{g^2 \nu^2(E_F ) k_B^3}{\hbar k_F \ell}. \nonumber	 
\end{align}
Here, $g$ is the electron-phonon coupling, $\nu(E_F)$ is the density of states at the Fermi level per valley/spin flavour, $k_F$ is the Fermi wavevector and $\ell$ is the mean free path. In high-quality samples and at cryogenic temperatures, the device size typically limits the latter. For low doping levels ($10^{12} \textrm{ cm}^{-2}$), $0.1<\ell<2$ $\mu$m and $T=25$ K, $\tau_{SC}=0.5-11 \textrm{ ns}$.

\textbf{Cooling due to lateral diffusion}
The lateral diffusion of photoexcited carriers reduces the hot electron temperature when the cooling length is greater than the laser spot size. This effect is particularly relevant in high-mobility samples, as the Wiedemann-Franz law relates electrical to thermal conductivity. \cite{Massicotte2021} At low lattice temperatures efficient heat conduction manifests in our experiments as a shorter cooling time. By considering the spatial evolution of a Gaussian heat spot induced by the laser pulse, \cite{Pogna2021a} we describe the temperature dynamics by:
\begin{equation*}
T_e(t)=2\pi A_{pu} A_{pr} \frac{\sigma_{pu}^2 \sigma_{pr}^2}{\sigma_{pu}^2+\sigma_{pr}^2+2Dt},
\end{equation*}
where $A$ and $\sigma$ are the peak intensity of the pump (\textit{pu}) and probe (\textit{pr}). Clearly, this effect is greater for smaller spot sizes and larger electronic heat diffusivities (\textit{D}).
Using a diffusivity of $D=750 \textrm{ cm}^2 \textrm{s}^{-1}$, and pump-probe spot sizes of $\sigma\approx0.9$ $\mu$m we find a cooling time of $\tau_{diff}\approx18\textrm{ ps}$. For $\sigma\approx1.4$ $\mu$m, $\tau_{diff}\approx45\textrm{ ps}$, see Fig. \ref{fig2}b. 

\textbf{Cooling rate at low temperatures}
The cooling rate in Fig.~\ref{fig3}f is estimated by $\frac{J(T_{el},T_{ph})}{C(T_{el})(T_{el}-T_{ph})}$, where $J$ is the cooling power, $C(T_{el})$ is the electron specific heat, and $T_{el}$ ($T_{ph}$) is the electron (phonon) temperature.
To evaluate $J$ and $C$, we consider an effective two-band model similar to pristine graphene used in Ref.~\cite{Ishizuka2021}.
Following the previous study, we use the electron-phonon interaction for the Wannier orbital radius $\xi=a/6$ where $a$ is the lattice parameter.
In the Boltzmann theory, the cooling power $J$ by electron-phonon scattering reads~\cite{Ishizuka2021}
\begin{align}
J=&\sum_{n,n'}J_{n,n'},\nonumber\\
J_{n,n'}=&\frac{2\pi}{V^2}\sum_{m,\vec k,\vec k'}\|g^{nn'}_{\overline{\vec k-\vec k'},m}\|^2\omega_{\overline{\vec k-\vec k'},m}^2N_{\overline{\vec k-\vec k'},m}\nonumber\\
&\times\left\{f_{\vec k'n'}[1-f_{\vec kn}]e^{\beta_\text{ph}\omega_{\overline{\vec k-\vec k'},m}}-f_{\vec kn}[1-f_{\vec k'n'}]\right\}\nonumber\\
&\times\delta(\varepsilon_{n'}-\varepsilon_{n}-\omega_{\overline{\vec k-\vec k'},m}),\label{eq:Jnn}
\end{align}
where $J_{n,n'}$ is the contribution from the scattering between $n$th and $n'$th bands, $V$ is the volume of the system, $g^{nn'}_{\overline{\vec k-\vec k'},m}$ is the coupling constant, $\varepsilon_{\vec kn}$ is the one-particle eigenenergy of the eigenstate in $n$th band with momentum $\vec k$, $\omega_{\vec qm}$ is the phonon eigenenergy in the $m$th band with momentum $\vec q$, and $\beta_{el}=1/k_B T_{el}$ ($\beta_{ph}=1/k_B T_{ph}$) is the inverse temperature of electrons (phonons) with $k_B$ being the Boltzmann constant, respectively; $f_{\vec kn}=\frac1{e^{\beta_\text{e}(\varepsilon_{\vec kn}-\mu)}+1}$ and $N_{\vec qm}=\frac1{e^{\beta_\text{ph}\omega_{\vec km}}-1}$ are respectively the Fermi and Bose distribution functions.
The estimation of specific heat uses the fluctuation formula
\begin{align}
&\quad C(T)=k_B \left[\langle \varepsilon_{n\vec k}^2\rangle -\frac{\langle \varepsilon_{n\vec k}\rangle^2}{\langle 1\rangle}\right],\label{eq:C}\\
&\langle O_{n\vec k}\rangle = \sum_n\int \frac{dk^d}{(2\pi)^d} \frac{\beta^2O_{n\vec k}}{4\cosh^2\left[\frac{\beta(\varepsilon_{n\vec k}-\mu)}2\right]}.
\end{align}

Note that the common formula for Fermi-degenerate electron systems does not apply here as the temperature exceeds the Fermi energy at $T\gtrsim 100$ K.
This model gives a good approximation when the temperature is much lower than the energy gap separating the flat band from high-energy dispersive bands.

\textbf{Cooling rate at high temperatures}
At high temperatures, we cannot neglect the high-energy bands because the electron temperature exceeds the band gap. In such a case, the Umklapp scattering involving high-energy phonons contributes to electron cooling due to a large number of high-energy phonons. Hence, we also expect that Umklapp scattering plays a key role in the high temperature regime.

To study the electron-lattice cooling involving the interband processes, we assume the electrons only couple to phonons with energies below a cutoff $\Lambda_{ph}$.
This assumption is justifiable in a system where the electron-phonon coupling between the electrons and the acoustic phonons reduces exponentially as the momentum increases.
In a system with compact Wannier orbitals, $\Lambda_{ph}$ becomes a few times higher than the energy of folded acoustic bands. 
Hence, a large $\Lambda_{ph}$, considerably larger than the phonon bandwidth of the folded acoustic phonons, represents the enhanced coupling by compact Wannier orbitals.
Below, we label the folded acoustic bands by an integer $m$ and define the high-temperature limit as $T_{el}>T_{ph}\gg\Lambda_{ph}$.

At high temperatures, the cooling power in Eq.~\eqref{eq:Jnn} reads
\begin{align*}
J_{nn'}=&\frac{\pi}V\sum_m\|g_m^{nn'}\|\omega_m^2\rho_n\rho_n'\left[T_{el}-T_{ph}\right]\times\nonumber\\
&\left[\tanh(\frac{\beta(b^m_{nn'}-\mu)}2)-\tanh(\frac{\beta(a^m_{nn'}-\mu)}2)\right],
\end{align*}
where $\rho_n$ is the density of states (DOS) for the $n$th band (we assume a constant DOS with the bandwidth $W_n$), and $a^m_{nn'}=\text{max}(\varepsilon_n^--\varepsilon_{n'}^--\omega_m)$ [$b^m_{nn'}=\text{min}(\varepsilon_n^+-\varepsilon_{n'}^+-\omega_m)$] with $\varepsilon_n^\pm$ being the energy of the top and bottom edge of the electron band.
Here, we approximated the phonon energy as $\omega_{n\vec k}\sim\omega_n$ considering the small Brillouin zone, and the coupling constant $g_m^{nn'}(\vec k)\sim g_m^{nn'}$ which is valid in the small orbital radius limit.

We apply the above formula to a four-band model consisting of two flat and two dispersive bands.
The two flat bands are at energies $0\le\varepsilon\le W$ and $-W\le\varepsilon\le 0$ with DOS $\rho_0$, and the two dispersive bands are $W<\Delta\le\varepsilon\le \Lambda$ and $-\Lambda\le\varepsilon\le -\Delta < -W$ with DOS $\rho_1$ (Fig.~\ref{fig4}b).
To the leading order in $T_{el}$, the cooling power reads
\begin{align*}
J=\pi\sum_m(\|g^{1,1}_m\|^2+\|g^{-1,-1}_m\|^2)\omega_m^2[T_{el}-T_{ph}]\rho_1^2.
\end{align*}
Hence, the cooling rate becomes $\tau^{-1}=\frac{6\rho_1}{\pi VT_{el}}\sum_m(\|g^{1,1}_m\|^2+\|g^{-1,-1}_m\|^2)\omega_m^2$, independent of phonon temperature, $T_{ph}$.

%\backmatter

\section{Supplementary information}

This article has an accompanying supplementary file.

\section{Acknowledgements}
We would like to thank Nick Feldman for his contribution to preliminary experiments. ICN2 was supported by the Severo Ochoa program from Spanish MINECO Grant No. SEV-2017-0706. 
R.L.M. acknowledges that this project has received funding from the “Secretaria d’Universitats I Recerca de la Generalitat de Catalunya, as well as the European Social Fund (L'FSE inverteix en el teu futur)—FEDER.
H.I. acknowledges support from  JSPS KAKENHI (Grant Numbers JP19K14649).
J.D.M. acknowledges support from the INphINIT ‘la Caixa’ Foundation (ID 100010434) fellowship programme (LCF/BQ/DI19/11730021). K.W. and T.T. acknowledge support from the JSPS KAKENHI (Grant Numbers 19H05790 and 20H00354 and 21H05233).
K.J.T. acknowledges funding from the European Union's Horizon 2020 research and innovation program under Grant Agreement No. 804349 (ERC StG CUHL), RYC fellowship No. RYC-2017-22330 and IAE project PID2019-111673GB-I00.

\bibliography{Mehew_MATBG_cooling_dynamics_v8.bib}% common bib file

\begin{thebibliography}{45}%
\makeatletter
\providecommand \@ifxundefined [1]{%
 \@ifx{#1\undefined}
}%
\providecommand \@ifnum [1]{%
 \ifnum #1\expandafter \@firstoftwo
 \else \expandafter \@secondoftwo
 \fi
}%
\providecommand \@ifx [1]{%
 \ifx #1\expandafter \@firstoftwo
 \else \expandafter \@secondoftwo
 \fi
}%
\providecommand \natexlab [1]{#1}%
\providecommand \enquote  [1]{``#1''}%
\providecommand \bibnamefont  [1]{#1}%
\providecommand \bibfnamefont [1]{#1}%
\providecommand \citenamefont [1]{#1}%
\providecommand \href@noop [0]{\@secondoftwo}%
\providecommand \href [0]{\begingroup \@sanitize@url \@href}%
\providecommand \@href[1]{\@@startlink{#1}\@@href}%
\providecommand \@@href[1]{\endgroup#1\@@endlink}%
\providecommand \@sanitize@url [0]{\catcode `\\12\catcode `\$12\catcode
  `\&12\catcode `\#12\catcode `\^12\catcode `\_12\catcode `\%12\relax}%
\providecommand \@@startlink[1]{}%
\providecommand \@@endlink[0]{}%
\providecommand \url  [0]{\begingroup\@sanitize@url \@url }%
\providecommand \@url [1]{\endgroup\@href {#1}{\urlprefix }}%
\providecommand \urlprefix  [0]{URL }%
\providecommand \Eprint [0]{\href }%
\providecommand \doibase [0]{https://doi.org/}%
\providecommand \selectlanguage [0]{\@gobble}%
\providecommand \bibinfo  [0]{\@secondoftwo}%
\providecommand \bibfield  [0]{\@secondoftwo}%
\providecommand \translation [1]{[#1]}%
\providecommand \BibitemOpen [0]{}%
\providecommand \bibitemStop [0]{}%
\providecommand \bibitemNoStop [0]{.\EOS\space}%
\providecommand \EOS [0]{\spacefactor3000\relax}%
\providecommand \BibitemShut  [1]{\csname bibitem#1\endcsname}%
\let\auto@bib@innerbib\@empty
%</preamble>
\bibitem [{\citenamefont {Cao}\ \emph {et~al.}(2018{\natexlab{a}})\citenamefont
  {Cao}, \citenamefont {Fatemi}, \citenamefont {Fang}, \citenamefont
  {Watanabe}, \citenamefont {Taniguchi}, \citenamefont {Kaxiras},\ and\
  \citenamefont {Jarillo-Herrero}}]{Cao2018}%
  \BibitemOpen
  \bibfield  {author} {\bibinfo {author} {\bibfnamefont {Y.}~\bibnamefont
  {Cao}}, \bibinfo {author} {\bibfnamefont {V.}~\bibnamefont {Fatemi}},
  \bibinfo {author} {\bibfnamefont {S.}~\bibnamefont {Fang}}, \bibinfo {author}
  {\bibfnamefont {K.}~\bibnamefont {Watanabe}}, \bibinfo {author}
  {\bibfnamefont {T.}~\bibnamefont {Taniguchi}}, \bibinfo {author}
  {\bibfnamefont {E.}~\bibnamefont {Kaxiras}},\ and\ \bibinfo {author}
  {\bibfnamefont {P.}~\bibnamefont {Jarillo-Herrero}},\ }\href
  {https://doi.org/10.1038/nature26160} {\bibfield  {journal} {\bibinfo
  {journal} {Nature}\ }\textbf {\bibinfo {volume} {556}},\ \bibinfo {pages}
  {43} (\bibinfo {year} {2018}{\natexlab{a}})}\BibitemShut {NoStop}%
\bibitem [{\citenamefont {Lu}\ \emph {et~al.}(2019)\citenamefont {Lu},
  \citenamefont {Stepanov}, \citenamefont {Yang}, \citenamefont {Xie},
  \citenamefont {Aamir}, \citenamefont {Das}, \citenamefont {Urgell},
  \citenamefont {Watanabe}, \citenamefont {Taniguchi}, \citenamefont {Zhang}
  \emph {et~al.}}]{Lu2019}%
  \BibitemOpen
  \bibfield  {author} {\bibinfo {author} {\bibfnamefont {X.}~\bibnamefont
  {Lu}}, \bibinfo {author} {\bibfnamefont {P.}~\bibnamefont {Stepanov}},
  \bibinfo {author} {\bibfnamefont {W.}~\bibnamefont {Yang}}, \bibinfo {author}
  {\bibfnamefont {M.}~\bibnamefont {Xie}}, \bibinfo {author} {\bibfnamefont
  {M.~A.}\ \bibnamefont {Aamir}}, \bibinfo {author} {\bibfnamefont
  {I.}~\bibnamefont {Das}}, \bibinfo {author} {\bibfnamefont {C.}~\bibnamefont
  {Urgell}}, \bibinfo {author} {\bibfnamefont {K.}~\bibnamefont {Watanabe}},
  \bibinfo {author} {\bibfnamefont {T.}~\bibnamefont {Taniguchi}}, \bibinfo
  {author} {\bibfnamefont {G.}~\bibnamefont {Zhang}}, \emph {et~al.},\
  }\href@noop {} {\bibfield  {journal} {\bibinfo  {journal} {Nature}\ }\textbf
  {\bibinfo {volume} {574}},\ \bibinfo {pages} {653} (\bibinfo {year}
  {2019})}\BibitemShut {NoStop}%
\bibitem [{\citenamefont {Yankowitz}\ \emph {et~al.}(2019)\citenamefont
  {Yankowitz}, \citenamefont {Chen}, \citenamefont {Polshyn}, \citenamefont
  {Zhang}, \citenamefont {Watanabe}, \citenamefont {Taniguchi}, \citenamefont
  {Graf}, \citenamefont {Young},\ and\ \citenamefont {Dean}}]{Yankowitz2019}%
  \BibitemOpen
  \bibfield  {author} {\bibinfo {author} {\bibfnamefont {M.}~\bibnamefont
  {Yankowitz}}, \bibinfo {author} {\bibfnamefont {S.}~\bibnamefont {Chen}},
  \bibinfo {author} {\bibfnamefont {H.}~\bibnamefont {Polshyn}}, \bibinfo
  {author} {\bibfnamefont {Y.}~\bibnamefont {Zhang}}, \bibinfo {author}
  {\bibfnamefont {K.}~\bibnamefont {Watanabe}}, \bibinfo {author}
  {\bibfnamefont {T.}~\bibnamefont {Taniguchi}}, \bibinfo {author}
  {\bibfnamefont {D.}~\bibnamefont {Graf}}, \bibinfo {author} {\bibfnamefont
  {A.~F.}\ \bibnamefont {Young}},\ and\ \bibinfo {author} {\bibfnamefont
  {C.~R.}\ \bibnamefont {Dean}},\ }\href@noop {} {\bibfield  {journal}
  {\bibinfo  {journal} {Science}\ }\textbf {\bibinfo {volume} {363}},\ \bibinfo
  {pages} {1059} (\bibinfo {year} {2019})}\BibitemShut {NoStop}%
\bibitem [{\citenamefont {Stepanov}\ \emph {et~al.}(2020)\citenamefont
  {Stepanov}, \citenamefont {Das}, \citenamefont {Lu}, \citenamefont
  {Fahimniya}, \citenamefont {Watanabe}, \citenamefont {Taniguchi},
  \citenamefont {Koppens}, \citenamefont {Lischner}, \citenamefont {Levitov},\
  and\ \citenamefont {Efetov}}]{Stepanov2020}%
  \BibitemOpen
  \bibfield  {author} {\bibinfo {author} {\bibfnamefont {P.}~\bibnamefont
  {Stepanov}}, \bibinfo {author} {\bibfnamefont {I.}~\bibnamefont {Das}},
  \bibinfo {author} {\bibfnamefont {X.}~\bibnamefont {Lu}}, \bibinfo {author}
  {\bibfnamefont {A.}~\bibnamefont {Fahimniya}}, \bibinfo {author}
  {\bibfnamefont {K.}~\bibnamefont {Watanabe}}, \bibinfo {author}
  {\bibfnamefont {T.}~\bibnamefont {Taniguchi}}, \bibinfo {author}
  {\bibfnamefont {F.~H.}\ \bibnamefont {Koppens}}, \bibinfo {author}
  {\bibfnamefont {J.}~\bibnamefont {Lischner}}, \bibinfo {author}
  {\bibfnamefont {L.}~\bibnamefont {Levitov}},\ and\ \bibinfo {author}
  {\bibfnamefont {D.~K.}\ \bibnamefont {Efetov}},\ }\href@noop {} {\bibfield
  {journal} {\bibinfo  {journal} {Nature}\ }\textbf {\bibinfo {volume} {583}},\
  \bibinfo {pages} {375} (\bibinfo {year} {2020})}\BibitemShut {NoStop}%
\bibitem [{\citenamefont {Cao}\ \emph {et~al.}(2018{\natexlab{b}})\citenamefont
  {Cao}, \citenamefont {Fatemi}, \citenamefont {Demir}, \citenamefont {Fang},
  \citenamefont {Tomarken}, \citenamefont {Luo}, \citenamefont
  {Sanchez-Yamagishi}, \citenamefont {Watanabe}, \citenamefont {Taniguchi},
  \citenamefont {Kaxiras}, \citenamefont {Ashoori},\ and\ \citenamefont
  {Jarillo-Herrero}}]{Cao2018b}%
  \BibitemOpen
  \bibfield  {author} {\bibinfo {author} {\bibfnamefont {Y.}~\bibnamefont
  {Cao}}, \bibinfo {author} {\bibfnamefont {V.}~\bibnamefont {Fatemi}},
  \bibinfo {author} {\bibfnamefont {A.}~\bibnamefont {Demir}}, \bibinfo
  {author} {\bibfnamefont {S.}~\bibnamefont {Fang}}, \bibinfo {author}
  {\bibfnamefont {S.~L.}\ \bibnamefont {Tomarken}}, \bibinfo {author}
  {\bibfnamefont {J.~Y.}\ \bibnamefont {Luo}}, \bibinfo {author} {\bibfnamefont
  {J.~D.}\ \bibnamefont {Sanchez-Yamagishi}}, \bibinfo {author} {\bibfnamefont
  {K.}~\bibnamefont {Watanabe}}, \bibinfo {author} {\bibfnamefont
  {T.}~\bibnamefont {Taniguchi}}, \bibinfo {author} {\bibfnamefont
  {E.}~\bibnamefont {Kaxiras}}, \bibinfo {author} {\bibfnamefont {R.~C.}\
  \bibnamefont {Ashoori}},\ and\ \bibinfo {author} {\bibfnamefont
  {P.}~\bibnamefont {Jarillo-Herrero}},\ }\href
  {https://doi.org/10.1038/nature26154} {\bibfield  {journal} {\bibinfo
  {journal} {Nature}\ }\textbf {\bibinfo {volume} {556}},\ \bibinfo {pages}
  {80} (\bibinfo {year} {2018}{\natexlab{b}})}\BibitemShut {NoStop}%
\bibitem [{\citenamefont {Wong}\ \emph {et~al.}(2020)\citenamefont {Wong},
  \citenamefont {Nuckolls}, \citenamefont {Oh}, \citenamefont {Lian},
  \citenamefont {Xie}, \citenamefont {Jeon}, \citenamefont {Watanabe},
  \citenamefont {Taniguchi}, \citenamefont {Bernevig},\ and\ \citenamefont
  {Yazdani}}]{Wong2020}%
  \BibitemOpen
  \bibfield  {author} {\bibinfo {author} {\bibfnamefont {D.}~\bibnamefont
  {Wong}}, \bibinfo {author} {\bibfnamefont {K.~P.}\ \bibnamefont {Nuckolls}},
  \bibinfo {author} {\bibfnamefont {M.}~\bibnamefont {Oh}}, \bibinfo {author}
  {\bibfnamefont {B.}~\bibnamefont {Lian}}, \bibinfo {author} {\bibfnamefont
  {Y.}~\bibnamefont {Xie}}, \bibinfo {author} {\bibfnamefont {S.}~\bibnamefont
  {Jeon}}, \bibinfo {author} {\bibfnamefont {K.}~\bibnamefont {Watanabe}},
  \bibinfo {author} {\bibfnamefont {T.}~\bibnamefont {Taniguchi}}, \bibinfo
  {author} {\bibfnamefont {B.~A.}\ \bibnamefont {Bernevig}},\ and\ \bibinfo
  {author} {\bibfnamefont {A.}~\bibnamefont {Yazdani}},\ }\href@noop {}
  {\bibfield  {journal} {\bibinfo  {journal} {Nature}\ }\textbf {\bibinfo
  {volume} {582}},\ \bibinfo {pages} {198} (\bibinfo {year}
  {2020})}\BibitemShut {NoStop}%
\bibitem [{\citenamefont {Sharpe}\ \emph {et~al.}(2019)\citenamefont {Sharpe},
  \citenamefont {Fox}, \citenamefont {Barnard}, \citenamefont {Finney},
  \citenamefont {Watanabe}, \citenamefont {Taniguchi}, \citenamefont
  {Kastner},\ and\ \citenamefont {Goldhaber-Gordon}}]{Sharpe2019}%
  \BibitemOpen
  \bibfield  {author} {\bibinfo {author} {\bibfnamefont {A.~L.}\ \bibnamefont
  {Sharpe}}, \bibinfo {author} {\bibfnamefont {E.~J.}\ \bibnamefont {Fox}},
  \bibinfo {author} {\bibfnamefont {A.~W.}\ \bibnamefont {Barnard}}, \bibinfo
  {author} {\bibfnamefont {J.}~\bibnamefont {Finney}}, \bibinfo {author}
  {\bibfnamefont {K.}~\bibnamefont {Watanabe}}, \bibinfo {author}
  {\bibfnamefont {T.}~\bibnamefont {Taniguchi}}, \bibinfo {author}
  {\bibfnamefont {M.}~\bibnamefont {Kastner}},\ and\ \bibinfo {author}
  {\bibfnamefont {D.}~\bibnamefont {Goldhaber-Gordon}},\ }\href@noop {}
  {\bibfield  {journal} {\bibinfo  {journal} {Science}\ }\textbf {\bibinfo
  {volume} {365}},\ \bibinfo {pages} {605} (\bibinfo {year}
  {2019})}\BibitemShut {NoStop}%
\bibitem [{\citenamefont {Regan}\ \emph {et~al.}(2020)\citenamefont {Regan},
  \citenamefont {Wang}, \citenamefont {Jin}, \citenamefont {Bakti~Utama},
  \citenamefont {Gao}, \citenamefont {Wei}, \citenamefont {Zhao}, \citenamefont
  {Zhao}, \citenamefont {Zhang}, \citenamefont {Yumigeta} \emph
  {et~al.}}]{Regan2020}%
  \BibitemOpen
  \bibfield  {author} {\bibinfo {author} {\bibfnamefont {E.~C.}\ \bibnamefont
  {Regan}}, \bibinfo {author} {\bibfnamefont {D.}~\bibnamefont {Wang}},
  \bibinfo {author} {\bibfnamefont {C.}~\bibnamefont {Jin}}, \bibinfo {author}
  {\bibfnamefont {M.~I.}\ \bibnamefont {Bakti~Utama}}, \bibinfo {author}
  {\bibfnamefont {B.}~\bibnamefont {Gao}}, \bibinfo {author} {\bibfnamefont
  {X.}~\bibnamefont {Wei}}, \bibinfo {author} {\bibfnamefont {S.}~\bibnamefont
  {Zhao}}, \bibinfo {author} {\bibfnamefont {W.}~\bibnamefont {Zhao}}, \bibinfo
  {author} {\bibfnamefont {Z.}~\bibnamefont {Zhang}}, \bibinfo {author}
  {\bibfnamefont {K.}~\bibnamefont {Yumigeta}}, \emph {et~al.},\ }\href@noop {}
  {\bibfield  {journal} {\bibinfo  {journal} {Nature}\ }\textbf {\bibinfo
  {volume} {579}},\ \bibinfo {pages} {359} (\bibinfo {year}
  {2020})}\BibitemShut {NoStop}%
\bibitem [{\citenamefont {Tang}\ \emph {et~al.}(2020)\citenamefont {Tang},
  \citenamefont {Li}, \citenamefont {Li}, \citenamefont {Xu}, \citenamefont
  {Liu}, \citenamefont {Barmak}, \citenamefont {Watanabe}, \citenamefont
  {Taniguchi}, \citenamefont {MacDonald}, \citenamefont {Shan} \emph
  {et~al.}}]{Tang2020}%
  \BibitemOpen
  \bibfield  {author} {\bibinfo {author} {\bibfnamefont {Y.}~\bibnamefont
  {Tang}}, \bibinfo {author} {\bibfnamefont {L.}~\bibnamefont {Li}}, \bibinfo
  {author} {\bibfnamefont {T.}~\bibnamefont {Li}}, \bibinfo {author}
  {\bibfnamefont {Y.}~\bibnamefont {Xu}}, \bibinfo {author} {\bibfnamefont
  {S.}~\bibnamefont {Liu}}, \bibinfo {author} {\bibfnamefont {K.}~\bibnamefont
  {Barmak}}, \bibinfo {author} {\bibfnamefont {K.}~\bibnamefont {Watanabe}},
  \bibinfo {author} {\bibfnamefont {T.}~\bibnamefont {Taniguchi}}, \bibinfo
  {author} {\bibfnamefont {A.~H.}\ \bibnamefont {MacDonald}}, \bibinfo {author}
  {\bibfnamefont {J.}~\bibnamefont {Shan}}, \emph {et~al.},\ }\href@noop {}
  {\bibfield  {journal} {\bibinfo  {journal} {Nature}\ }\textbf {\bibinfo
  {volume} {579}},\ \bibinfo {pages} {353} (\bibinfo {year}
  {2020})}\BibitemShut {NoStop}%
\bibitem [{\citenamefont {Wang}\ \emph {et~al.}(2022)\citenamefont {Wang},
  \citenamefont {Xiao}, \citenamefont {Park}, \citenamefont {Zhu},
  \citenamefont {Wang}, \citenamefont {Taniguchi}, \citenamefont {Watanabe},
  \citenamefont {Yan}, \citenamefont {Xiao}, \citenamefont {Gamelin} \emph
  {et~al.}}]{Wang2022}%
  \BibitemOpen
  \bibfield  {author} {\bibinfo {author} {\bibfnamefont {X.}~\bibnamefont
  {Wang}}, \bibinfo {author} {\bibfnamefont {C.}~\bibnamefont {Xiao}}, \bibinfo
  {author} {\bibfnamefont {H.}~\bibnamefont {Park}}, \bibinfo {author}
  {\bibfnamefont {J.}~\bibnamefont {Zhu}}, \bibinfo {author} {\bibfnamefont
  {C.}~\bibnamefont {Wang}}, \bibinfo {author} {\bibfnamefont {T.}~\bibnamefont
  {Taniguchi}}, \bibinfo {author} {\bibfnamefont {K.}~\bibnamefont {Watanabe}},
  \bibinfo {author} {\bibfnamefont {J.}~\bibnamefont {Yan}}, \bibinfo {author}
  {\bibfnamefont {D.}~\bibnamefont {Xiao}}, \bibinfo {author} {\bibfnamefont
  {D.~R.}\ \bibnamefont {Gamelin}}, \emph {et~al.},\ }\href@noop {} {\bibfield
  {journal} {\bibinfo  {journal} {Nature}\ }\textbf {\bibinfo {volume} {604}},\
  \bibinfo {pages} {468} (\bibinfo {year} {2022})}\BibitemShut {NoStop}%
\bibitem [{\citenamefont {Seyler}\ \emph {et~al.}(2019)\citenamefont {Seyler},
  \citenamefont {Rivera}, \citenamefont {Yu}, \citenamefont {Wilson},
  \citenamefont {Ray}, \citenamefont {Mandrus}, \citenamefont {Yan},
  \citenamefont {Yao},\ and\ \citenamefont {Xu}}]{Seyler2019}%
  \BibitemOpen
  \bibfield  {author} {\bibinfo {author} {\bibfnamefont {K.~L.}\ \bibnamefont
  {Seyler}}, \bibinfo {author} {\bibfnamefont {P.}~\bibnamefont {Rivera}},
  \bibinfo {author} {\bibfnamefont {H.}~\bibnamefont {Yu}}, \bibinfo {author}
  {\bibfnamefont {N.~P.}\ \bibnamefont {Wilson}}, \bibinfo {author}
  {\bibfnamefont {E.~L.}\ \bibnamefont {Ray}}, \bibinfo {author} {\bibfnamefont
  {D.~G.}\ \bibnamefont {Mandrus}}, \bibinfo {author} {\bibfnamefont
  {J.}~\bibnamefont {Yan}}, \bibinfo {author} {\bibfnamefont {W.}~\bibnamefont
  {Yao}},\ and\ \bibinfo {author} {\bibfnamefont {X.}~\bibnamefont {Xu}},\
  }\href@noop {} {\bibfield  {journal} {\bibinfo  {journal} {Nature}\ }\textbf
  {\bibinfo {volume} {567}},\ \bibinfo {pages} {66} (\bibinfo {year}
  {2019})}\BibitemShut {NoStop}%
\bibitem [{\citenamefont {Alexeev}\ \emph {et~al.}(2019)\citenamefont
  {Alexeev}, \citenamefont {Ruiz-Tijerina}, \citenamefont {Danovich},
  \citenamefont {Hamer}, \citenamefont {Terry}, \citenamefont {Nayak},
  \citenamefont {Ahn}, \citenamefont {Pak}, \citenamefont {Lee}, \citenamefont
  {Sohn} \emph {et~al.}}]{Alexeev2019}%
  \BibitemOpen
  \bibfield  {author} {\bibinfo {author} {\bibfnamefont {E.~M.}\ \bibnamefont
  {Alexeev}}, \bibinfo {author} {\bibfnamefont {D.~A.}\ \bibnamefont
  {Ruiz-Tijerina}}, \bibinfo {author} {\bibfnamefont {M.}~\bibnamefont
  {Danovich}}, \bibinfo {author} {\bibfnamefont {M.~J.}\ \bibnamefont {Hamer}},
  \bibinfo {author} {\bibfnamefont {D.~J.}\ \bibnamefont {Terry}}, \bibinfo
  {author} {\bibfnamefont {P.~K.}\ \bibnamefont {Nayak}}, \bibinfo {author}
  {\bibfnamefont {S.}~\bibnamefont {Ahn}}, \bibinfo {author} {\bibfnamefont
  {S.}~\bibnamefont {Pak}}, \bibinfo {author} {\bibfnamefont {J.}~\bibnamefont
  {Lee}}, \bibinfo {author} {\bibfnamefont {J.~I.}\ \bibnamefont {Sohn}}, \emph
  {et~al.},\ }\href@noop {} {\bibfield  {journal} {\bibinfo  {journal}
  {Nature}\ }\textbf {\bibinfo {volume} {567}},\ \bibinfo {pages} {81}
  (\bibinfo {year} {2019})}\BibitemShut {NoStop}%
\bibitem [{\citenamefont {Lin}\ \emph {et~al.}(2018)\citenamefont {Lin},
  \citenamefont {Tan}, \citenamefont {Wu}, \citenamefont {Chen}, \citenamefont
  {Wang}, \citenamefont {Pan}, \citenamefont {Zhang}, \citenamefont {Cong},
  \citenamefont {Zhang}, \citenamefont {Ji}, \citenamefont {Hu}, \citenamefont
  {Liu},\ and\ \citenamefont {Tan}}]{Lin2018}%
  \BibitemOpen
  \bibfield  {author} {\bibinfo {author} {\bibfnamefont {M.-L.}\ \bibnamefont
  {Lin}}, \bibinfo {author} {\bibfnamefont {Q.-H.}\ \bibnamefont {Tan}},
  \bibinfo {author} {\bibfnamefont {J.-B.}\ \bibnamefont {Wu}}, \bibinfo
  {author} {\bibfnamefont {X.-S.}\ \bibnamefont {Chen}}, \bibinfo {author}
  {\bibfnamefont {J.-H.}\ \bibnamefont {Wang}}, \bibinfo {author}
  {\bibfnamefont {Y.-H.}\ \bibnamefont {Pan}}, \bibinfo {author} {\bibfnamefont
  {X.}~\bibnamefont {Zhang}}, \bibinfo {author} {\bibfnamefont
  {X.}~\bibnamefont {Cong}}, \bibinfo {author} {\bibfnamefont {J.}~\bibnamefont
  {Zhang}}, \bibinfo {author} {\bibfnamefont {W.}~\bibnamefont {Ji}}, \bibinfo
  {author} {\bibfnamefont {P.-A.}\ \bibnamefont {Hu}}, \bibinfo {author}
  {\bibfnamefont {K.-H.}\ \bibnamefont {Liu}},\ and\ \bibinfo {author}
  {\bibfnamefont {P.-H.}\ \bibnamefont {Tan}},\ }\href
  {https://doi.org/10.1021/acsnano.8b05006} {\bibfield  {journal} {\bibinfo
  {journal} {ACS Nano}\ }\textbf {\bibinfo {volume} {12}},\ \bibinfo {pages}
  {8770} (\bibinfo {year} {2018})},\ \bibinfo {note} {pMID: 30086224},\ \Eprint
  {https://arxiv.org/abs/https://doi.org/10.1021/acsnano.8b05006}
  {https://doi.org/10.1021/acsnano.8b05006} \BibitemShut {NoStop}%
\bibitem [{\citenamefont {Quan}\ \emph {et~al.}(2021)\citenamefont {Quan},
  \citenamefont {Linhart}, \citenamefont {Lin}, \citenamefont {Lee},
  \citenamefont {Zhu}, \citenamefont {Wang}, \citenamefont {Hsu}, \citenamefont
  {Choi}, \citenamefont {Embley}, \citenamefont {Young} \emph
  {et~al.}}]{Quan2021}%
  \BibitemOpen
  \bibfield  {author} {\bibinfo {author} {\bibfnamefont {J.}~\bibnamefont
  {Quan}}, \bibinfo {author} {\bibfnamefont {L.}~\bibnamefont {Linhart}},
  \bibinfo {author} {\bibfnamefont {M.-L.}\ \bibnamefont {Lin}}, \bibinfo
  {author} {\bibfnamefont {D.}~\bibnamefont {Lee}}, \bibinfo {author}
  {\bibfnamefont {J.}~\bibnamefont {Zhu}}, \bibinfo {author} {\bibfnamefont
  {C.-Y.}\ \bibnamefont {Wang}}, \bibinfo {author} {\bibfnamefont {W.-T.}\
  \bibnamefont {Hsu}}, \bibinfo {author} {\bibfnamefont {J.}~\bibnamefont
  {Choi}}, \bibinfo {author} {\bibfnamefont {J.}~\bibnamefont {Embley}},
  \bibinfo {author} {\bibfnamefont {C.}~\bibnamefont {Young}}, \emph {et~al.},\
  }\href@noop {} {\bibfield  {journal} {\bibinfo  {journal} {Nature materials}\
  }\textbf {\bibinfo {volume} {20}},\ \bibinfo {pages} {1100} (\bibinfo {year}
  {2021})}\BibitemShut {NoStop}%
\bibitem [{\citenamefont {Koshino}\ and\ \citenamefont
  {Son}(2019)}]{Koshino2019}%
  \BibitemOpen
  \bibfield  {author} {\bibinfo {author} {\bibfnamefont {M.}~\bibnamefont
  {Koshino}}\ and\ \bibinfo {author} {\bibfnamefont {Y.-W.}\ \bibnamefont
  {Son}},\ }\href {https://doi.org/10.1103/PhysRevB.100.075416} {\bibfield
  {journal} {\bibinfo  {journal} {Phys. Rev. B}\ }\textbf {\bibinfo {volume}
  {100}},\ \bibinfo {pages} {075416} (\bibinfo {year} {2019})}\BibitemShut
  {NoStop}%
\bibitem [{\citenamefont {Wu}\ \emph {et~al.}(2018)\citenamefont {Wu},
  \citenamefont {MacDonald},\ and\ \citenamefont {Martin}}]{Wu2018}%
  \BibitemOpen
  \bibfield  {author} {\bibinfo {author} {\bibfnamefont {F.}~\bibnamefont
  {Wu}}, \bibinfo {author} {\bibfnamefont {A.~H.}\ \bibnamefont {MacDonald}},\
  and\ \bibinfo {author} {\bibfnamefont {I.}~\bibnamefont {Martin}},\ }\href
  {https://doi.org/10.1103/PhysRevLett.121.257001} {\bibfield  {journal}
  {\bibinfo  {journal} {Phys. Rev. Lett.}\ }\textbf {\bibinfo {volume} {121}},\
  \bibinfo {pages} {257001} (\bibinfo {year} {2018})}\BibitemShut {NoStop}%
\bibitem [{\citenamefont {Choi}\ and\ \citenamefont {Choi}(2018)}]{Choi2018}%
  \BibitemOpen
  \bibfield  {author} {\bibinfo {author} {\bibfnamefont {Y.~W.}\ \bibnamefont
  {Choi}}\ and\ \bibinfo {author} {\bibfnamefont {H.~J.}\ \bibnamefont
  {Choi}},\ }\bibfield  {journal} {\bibinfo  {journal} {PHYSICAL REVIEW B}\
  }\textbf {\bibinfo {volume} {98}},\ \href
  {https://doi.org/10.1103/PhysRevB.98.241412} {10.1103/PhysRevB.98.241412}
  (\bibinfo {year} {2018})\BibitemShut {NoStop}%
\bibitem [{\citenamefont {Koshino}\ and\ \citenamefont
  {Nam}(2020)}]{Koshino2020}%
  \BibitemOpen
  \bibfield  {author} {\bibinfo {author} {\bibfnamefont {M.}~\bibnamefont
  {Koshino}}\ and\ \bibinfo {author} {\bibfnamefont {N.~N.~T.}\ \bibnamefont
  {Nam}},\ }\bibfield  {journal} {\bibinfo  {journal} {PHYSICAL REVIEW B}\
  }\textbf {\bibinfo {volume} {101}},\ \href
  {https://doi.org/10.1103/PhysRevB.101.195425} {10.1103/PhysRevB.101.195425}
  (\bibinfo {year} {2020})\BibitemShut {NoStop}%
\bibitem [{\citenamefont {Tielrooij}\ \emph {et~al.}(2015)\citenamefont
  {Tielrooij}, \citenamefont {Piatkowski}, \citenamefont {Massicotte},
  \citenamefont {Woessner}, \citenamefont {Ma}, \citenamefont {Lee},
  \citenamefont {Myhro}, \citenamefont {Lau}, \citenamefont {Jarillo-Herrero},
  \citenamefont {van Hulst} \emph {et~al.}}]{Tielrooij2015}%
  \BibitemOpen
  \bibfield  {author} {\bibinfo {author} {\bibfnamefont {K.-J.}\ \bibnamefont
  {Tielrooij}}, \bibinfo {author} {\bibfnamefont {L.}~\bibnamefont
  {Piatkowski}}, \bibinfo {author} {\bibfnamefont {M.}~\bibnamefont
  {Massicotte}}, \bibinfo {author} {\bibfnamefont {A.}~\bibnamefont
  {Woessner}}, \bibinfo {author} {\bibfnamefont {Q.}~\bibnamefont {Ma}},
  \bibinfo {author} {\bibfnamefont {Y.}~\bibnamefont {Lee}}, \bibinfo {author}
  {\bibfnamefont {K.~S.}\ \bibnamefont {Myhro}}, \bibinfo {author}
  {\bibfnamefont {C.~N.}\ \bibnamefont {Lau}}, \bibinfo {author} {\bibfnamefont
  {P.}~\bibnamefont {Jarillo-Herrero}}, \bibinfo {author} {\bibfnamefont
  {N.~F.}\ \bibnamefont {van Hulst}}, \emph {et~al.},\ }\href@noop {}
  {\bibfield  {journal} {\bibinfo  {journal} {Nature nanotechnology}\ }\textbf
  {\bibinfo {volume} {10}},\ \bibinfo {pages} {437} (\bibinfo {year}
  {2015})}\BibitemShut {NoStop}%
\bibitem [{\citenamefont {Bistritzer}\ and\ \citenamefont
  {MacDonald}(2009)}]{Bistritzer2009}%
  \BibitemOpen
  \bibfield  {author} {\bibinfo {author} {\bibfnamefont {R.}~\bibnamefont
  {Bistritzer}}\ and\ \bibinfo {author} {\bibfnamefont {A.~H.}\ \bibnamefont
  {MacDonald}},\ }\href {https://doi.org/10.1103/PhysRevLett.102.206410}
  {\bibfield  {journal} {\bibinfo  {journal} {Phys. Rev. Lett.}\ }\textbf
  {\bibinfo {volume} {102}},\ \bibinfo {pages} {206410} (\bibinfo {year}
  {2009})}\BibitemShut {NoStop}%
\bibitem [{\citenamefont {Song}\ \emph {et~al.}(2012)\citenamefont {Song},
  \citenamefont {Reizer},\ and\ \citenamefont {Levitov}}]{Song2012}%
  \BibitemOpen
  \bibfield  {author} {\bibinfo {author} {\bibfnamefont {J.~C.~W.}\
  \bibnamefont {Song}}, \bibinfo {author} {\bibfnamefont {M.~Y.}\ \bibnamefont
  {Reizer}},\ and\ \bibinfo {author} {\bibfnamefont {L.~S.}\ \bibnamefont
  {Levitov}},\ }\bibfield  {journal} {\bibinfo  {journal} {PHYSICAL REVIEW
  LETTERS}\ }\textbf {\bibinfo {volume} {109}},\ \href
  {https://doi.org/10.1103/PhysRevLett.109.106602}
  {10.1103/PhysRevLett.109.106602} (\bibinfo {year} {2012})\BibitemShut
  {NoStop}%
\bibitem [{\citenamefont {Graham}\ \emph {et~al.}(2013)\citenamefont {Graham},
  \citenamefont {Shi}, \citenamefont {Ralph}, \citenamefont {Park},\ and\
  \citenamefont {McEuen}}]{Graham2013}%
  \BibitemOpen
  \bibfield  {author} {\bibinfo {author} {\bibfnamefont {M.~W.}\ \bibnamefont
  {Graham}}, \bibinfo {author} {\bibfnamefont {S.-F.}\ \bibnamefont {Shi}},
  \bibinfo {author} {\bibfnamefont {D.~C.}\ \bibnamefont {Ralph}}, \bibinfo
  {author} {\bibfnamefont {J.}~\bibnamefont {Park}},\ and\ \bibinfo {author}
  {\bibfnamefont {P.~L.}\ \bibnamefont {McEuen}},\ }\href
  {https://doi.org/10.1038/nphys2493} {\bibfield  {journal} {\bibinfo
  {journal} {NATURE PHYSICS}\ }\textbf {\bibinfo {volume} {9}},\ \bibinfo
  {pages} {103} (\bibinfo {year} {2013})}\BibitemShut {NoStop}%
\bibitem [{\citenamefont {Kong}\ \emph {et~al.}(2018)\citenamefont {Kong},
  \citenamefont {Levitov}, \citenamefont {Halbertal},\ and\ \citenamefont
  {Zeldov}}]{Kong2018}%
  \BibitemOpen
  \bibfield  {author} {\bibinfo {author} {\bibfnamefont {J.~F.}\ \bibnamefont
  {Kong}}, \bibinfo {author} {\bibfnamefont {L.}~\bibnamefont {Levitov}},
  \bibinfo {author} {\bibfnamefont {D.}~\bibnamefont {Halbertal}},\ and\
  \bibinfo {author} {\bibfnamefont {E.}~\bibnamefont {Zeldov}},\ }\href
  {https://doi.org/10.1103/PhysRevB.97.245416} {\bibfield  {journal} {\bibinfo
  {journal} {Phys. Rev. B}\ }\textbf {\bibinfo {volume} {97}},\ \bibinfo
  {pages} {245416} (\bibinfo {year} {2018})}\BibitemShut {NoStop}%
\bibitem [{\citenamefont {Tielrooij}\ \emph {et~al.}(2018)\citenamefont
  {Tielrooij}, \citenamefont {Hesp}, \citenamefont {Principi}, \citenamefont
  {Lundeberg}, \citenamefont {Pogna}, \citenamefont {Banszerus}, \citenamefont
  {Mics}, \citenamefont {Massicotte}, \citenamefont {Schmidt}, \citenamefont
  {Davydovskaya}, \citenamefont {Purdie}, \citenamefont {Goykhman},
  \citenamefont {Soavi}, \citenamefont {Lombardo}, \citenamefont {Watanabe},
  \citenamefont {Taniguchi}, \citenamefont {Bonn}, \citenamefont
  {Turchinovich}, \citenamefont {Stampfer}, \citenamefont {Ferrari},
  \citenamefont {Cerullo}, \citenamefont {Polini},\ and\ \citenamefont
  {Koppens}}]{Tielrooij2018}%
  \BibitemOpen
  \bibfield  {author} {\bibinfo {author} {\bibfnamefont {K.-J.}\ \bibnamefont
  {Tielrooij}}, \bibinfo {author} {\bibfnamefont {N.~C.~H.}\ \bibnamefont
  {Hesp}}, \bibinfo {author} {\bibfnamefont {A.}~\bibnamefont {Principi}},
  \bibinfo {author} {\bibfnamefont {M.~B.}\ \bibnamefont {Lundeberg}}, \bibinfo
  {author} {\bibfnamefont {E.~A.~A.}\ \bibnamefont {Pogna}}, \bibinfo {author}
  {\bibfnamefont {L.}~\bibnamefont {Banszerus}}, \bibinfo {author}
  {\bibfnamefont {Z.}~\bibnamefont {Mics}}, \bibinfo {author} {\bibfnamefont
  {M.}~\bibnamefont {Massicotte}}, \bibinfo {author} {\bibfnamefont
  {P.}~\bibnamefont {Schmidt}}, \bibinfo {author} {\bibfnamefont
  {D.}~\bibnamefont {Davydovskaya}}, \bibinfo {author} {\bibfnamefont {D.~G.}\
  \bibnamefont {Purdie}}, \bibinfo {author} {\bibfnamefont {I.}~\bibnamefont
  {Goykhman}}, \bibinfo {author} {\bibfnamefont {G.}~\bibnamefont {Soavi}},
  \bibinfo {author} {\bibfnamefont {A.}~\bibnamefont {Lombardo}}, \bibinfo
  {author} {\bibfnamefont {K.}~\bibnamefont {Watanabe}}, \bibinfo {author}
  {\bibfnamefont {T.}~\bibnamefont {Taniguchi}}, \bibinfo {author}
  {\bibfnamefont {M.}~\bibnamefont {Bonn}}, \bibinfo {author} {\bibfnamefont
  {D.}~\bibnamefont {Turchinovich}}, \bibinfo {author} {\bibfnamefont
  {C.}~\bibnamefont {Stampfer}}, \bibinfo {author} {\bibfnamefont {A.~C.}\
  \bibnamefont {Ferrari}}, \bibinfo {author} {\bibfnamefont {G.}~\bibnamefont
  {Cerullo}}, \bibinfo {author} {\bibfnamefont {M.}~\bibnamefont {Polini}},\
  and\ \bibinfo {author} {\bibfnamefont {F.~H.~L.}\ \bibnamefont {Koppens}},\
  }\href {https://doi.org/10.1038/s41565-017-0008-8} {\bibfield  {journal}
  {\bibinfo  {journal} {NATURE NANOTECHNOLOGY}\ }\textbf {\bibinfo {volume}
  {13}},\ \bibinfo {pages} {41+} (\bibinfo {year} {2018})}\BibitemShut
  {NoStop}%
\bibitem [{\citenamefont {Massicotte}\ \emph {et~al.}(2021)\citenamefont
  {Massicotte}, \citenamefont {Soavi}, \citenamefont {Principi},\ and\
  \citenamefont {Tielrooij}}]{Massicotte2021}%
  \BibitemOpen
  \bibfield  {author} {\bibinfo {author} {\bibfnamefont {M.}~\bibnamefont
  {Massicotte}}, \bibinfo {author} {\bibfnamefont {G.}~\bibnamefont {Soavi}},
  \bibinfo {author} {\bibfnamefont {A.}~\bibnamefont {Principi}},\ and\
  \bibinfo {author} {\bibfnamefont {K.-J.}\ \bibnamefont {Tielrooij}},\ }\href
  {https://doi.org/10.1039/D0NR09166A} {\bibfield  {journal} {\bibinfo
  {journal} {Nanoscale}\ }\textbf {\bibinfo {volume} {13}},\ \bibinfo {pages}
  {8376} (\bibinfo {year} {2021})}\BibitemShut {NoStop}%
\bibitem [{\citenamefont {Pogna}\ \emph {et~al.}(2021)\citenamefont {Pogna},
  \citenamefont {Jia}, \citenamefont {Principi}, \citenamefont {Block},
  \citenamefont {Banszerus}, \citenamefont {Zhang}, \citenamefont {Liu},
  \citenamefont {Sohier}, \citenamefont {Forti}, \citenamefont
  {Soundarapandian}, \citenamefont {Terres}, \citenamefont {Mehew},
  \citenamefont {Trovatello}, \citenamefont {Coletti}, \citenamefont {Koppens},
  \citenamefont {Bonn}, \citenamefont {Wang}, \citenamefont {van Hulst},
  \citenamefont {Verstraete}, \citenamefont {Peng}, \citenamefont {Liu},
  \citenamefont {Stampfer}, \citenamefont {Cerullo},\ and\ \citenamefont
  {Tielrooij}}]{Pogna2021a}%
  \BibitemOpen
  \bibfield  {author} {\bibinfo {author} {\bibfnamefont {E.~A.~A.}\
  \bibnamefont {Pogna}}, \bibinfo {author} {\bibfnamefont {X.}~\bibnamefont
  {Jia}}, \bibinfo {author} {\bibfnamefont {A.}~\bibnamefont {Principi}},
  \bibinfo {author} {\bibfnamefont {A.}~\bibnamefont {Block}}, \bibinfo
  {author} {\bibfnamefont {L.}~\bibnamefont {Banszerus}}, \bibinfo {author}
  {\bibfnamefont {J.}~\bibnamefont {Zhang}}, \bibinfo {author} {\bibfnamefont
  {X.}~\bibnamefont {Liu}}, \bibinfo {author} {\bibfnamefont {T.}~\bibnamefont
  {Sohier}}, \bibinfo {author} {\bibfnamefont {S.}~\bibnamefont {Forti}},
  \bibinfo {author} {\bibfnamefont {K.}~\bibnamefont {Soundarapandian}},
  \bibinfo {author} {\bibfnamefont {B.}~\bibnamefont {Terres}}, \bibinfo
  {author} {\bibfnamefont {J.~D.}\ \bibnamefont {Mehew}}, \bibinfo {author}
  {\bibfnamefont {C.}~\bibnamefont {Trovatello}}, \bibinfo {author}
  {\bibfnamefont {C.}~\bibnamefont {Coletti}}, \bibinfo {author} {\bibfnamefont
  {F.~H.~L.}\ \bibnamefont {Koppens}}, \bibinfo {author} {\bibfnamefont
  {M.}~\bibnamefont {Bonn}}, \bibinfo {author} {\bibfnamefont {H.}~\bibnamefont
  {Wang}, \bibfnamefont {I}}, \bibinfo {author} {\bibfnamefont
  {N.}~\bibnamefont {van Hulst}}, \bibinfo {author} {\bibfnamefont {M.~J.}\
  \bibnamefont {Verstraete}}, \bibinfo {author} {\bibfnamefont
  {H.}~\bibnamefont {Peng}}, \bibinfo {author} {\bibfnamefont {Z.}~\bibnamefont
  {Liu}}, \bibinfo {author} {\bibfnamefont {C.}~\bibnamefont {Stampfer}},
  \bibinfo {author} {\bibfnamefont {G.}~\bibnamefont {Cerullo}},\ and\ \bibinfo
  {author} {\bibfnamefont {K.-J.}\ \bibnamefont {Tielrooij}},\ }\href
  {https://doi.org/10.1021/acsnano.0c10864} {\bibfield  {journal} {\bibinfo
  {journal} {ACS NANO}\ }\textbf {\bibinfo {volume} {15}},\ \bibinfo {pages}
  {11285} (\bibinfo {year} {2021})}\BibitemShut {NoStop}%
\bibitem [{\citenamefont {Kim}\ \emph {et~al.}(2021)\citenamefont {Kim},
  \citenamefont {Kim}, \citenamefont {Jha}, \citenamefont {Brar},\ and\
  \citenamefont {Atwater}}]{Kim2021}%
  \BibitemOpen
  \bibfield  {author} {\bibinfo {author} {\bibfnamefont {L.}~\bibnamefont
  {Kim}}, \bibinfo {author} {\bibfnamefont {S.}~\bibnamefont {Kim}}, \bibinfo
  {author} {\bibfnamefont {P.~K.}\ \bibnamefont {Jha}}, \bibinfo {author}
  {\bibfnamefont {V.~W.}\ \bibnamefont {Brar}},\ and\ \bibinfo {author}
  {\bibfnamefont {H.~A.}\ \bibnamefont {Atwater}},\ }\href
  {https://doi.org/10.1038/s41563-021-00935-2} {\bibfield  {journal} {\bibinfo
  {journal} {NATURE MATERIALS}\ }\textbf {\bibinfo {volume} {20}},\ \bibinfo
  {pages} {805+} (\bibinfo {year} {2021})}\BibitemShut {NoStop}%
\bibitem [{\citenamefont {Patel}\ \emph {et~al.}(2015)\citenamefont {Patel},
  \citenamefont {Havener}, \citenamefont {Brown}, \citenamefont {Liang},
  \citenamefont {Yang}, \citenamefont {Park},\ and\ \citenamefont
  {Graham}}]{Patel2015}%
  \BibitemOpen
  \bibfield  {author} {\bibinfo {author} {\bibfnamefont {H.}~\bibnamefont
  {Patel}}, \bibinfo {author} {\bibfnamefont {R.~W.}\ \bibnamefont {Havener}},
  \bibinfo {author} {\bibfnamefont {L.}~\bibnamefont {Brown}}, \bibinfo
  {author} {\bibfnamefont {Y.}~\bibnamefont {Liang}}, \bibinfo {author}
  {\bibfnamefont {L.}~\bibnamefont {Yang}}, \bibinfo {author} {\bibfnamefont
  {J.}~\bibnamefont {Park}},\ and\ \bibinfo {author} {\bibfnamefont {M.~W.}\
  \bibnamefont {Graham}},\ }\href
  {https://doi.org/10.1021/acs.nanolett.5b02035} {\bibfield  {journal}
  {\bibinfo  {journal} {NANO LETTERS}\ }\textbf {\bibinfo {volume} {15}},\
  \bibinfo {pages} {5932} (\bibinfo {year} {2015})}\BibitemShut {NoStop}%
\bibitem [{\citenamefont {Patel}\ \emph {et~al.}(2019)\citenamefont {Patel},
  \citenamefont {Huang}, \citenamefont {Kim}, \citenamefont {Park},\ and\
  \citenamefont {Graham}}]{Patel2019}%
  \BibitemOpen
  \bibfield  {author} {\bibinfo {author} {\bibfnamefont {H.}~\bibnamefont
  {Patel}}, \bibinfo {author} {\bibfnamefont {L.}~\bibnamefont {Huang}},
  \bibinfo {author} {\bibfnamefont {C.-J.}\ \bibnamefont {Kim}}, \bibinfo
  {author} {\bibfnamefont {J.}~\bibnamefont {Park}},\ and\ \bibinfo {author}
  {\bibfnamefont {M.~W.}\ \bibnamefont {Graham}},\ }\bibfield  {journal}
  {\bibinfo  {journal} {NATURE COMMUNICATIONS}\ }\textbf {\bibinfo {volume}
  {10}},\ \href {https://doi.org/10.1038/s41467-019-09097-x}
  {10.1038/s41467-019-09097-x} (\bibinfo {year} {2019})\BibitemShut {NoStop}%
\bibitem [{\citenamefont {Gadelha}\ \emph {et~al.}(2021)\citenamefont
  {Gadelha}, \citenamefont {Ohlberg}, \citenamefont {Rabelo}, \citenamefont
  {Neto}, \citenamefont {Vasconcelos}, \citenamefont {Campos}, \citenamefont
  {Lemos}, \citenamefont {Ornelas}, \citenamefont {Miranda}, \citenamefont
  {Nadas}, \citenamefont {Santana}, \citenamefont {Watanabe}, \citenamefont
  {Taniguchi}, \citenamefont {van Troeye}, \citenamefont {Lamparski},
  \citenamefont {Meunier}, \citenamefont {Nguyen}, \citenamefont {Paszko},
  \citenamefont {Charlier}, \citenamefont {Campos}, \citenamefont
  {Can{\c{c}}ado}, \citenamefont {Medeiros-Ribeiro},\ and\ \citenamefont
  {Jorio}}]{Gadelha2021}%
  \BibitemOpen
  \bibfield  {author} {\bibinfo {author} {\bibfnamefont {A.~C.}\ \bibnamefont
  {Gadelha}}, \bibinfo {author} {\bibfnamefont {D.~A.~A.}\ \bibnamefont
  {Ohlberg}}, \bibinfo {author} {\bibfnamefont {C.}~\bibnamefont {Rabelo}},
  \bibinfo {author} {\bibfnamefont {E.~G.~S.}\ \bibnamefont {Neto}}, \bibinfo
  {author} {\bibfnamefont {T.~L.}\ \bibnamefont {Vasconcelos}}, \bibinfo
  {author} {\bibfnamefont {J.~L.}\ \bibnamefont {Campos}}, \bibinfo {author}
  {\bibfnamefont {J.~S.}\ \bibnamefont {Lemos}}, \bibinfo {author}
  {\bibfnamefont {V.}~\bibnamefont {Ornelas}}, \bibinfo {author} {\bibfnamefont
  {D.}~\bibnamefont {Miranda}}, \bibinfo {author} {\bibfnamefont
  {R.}~\bibnamefont {Nadas}}, \bibinfo {author} {\bibfnamefont {F.~C.}\
  \bibnamefont {Santana}}, \bibinfo {author} {\bibfnamefont {K.}~\bibnamefont
  {Watanabe}}, \bibinfo {author} {\bibfnamefont {T.}~\bibnamefont {Taniguchi}},
  \bibinfo {author} {\bibfnamefont {B.}~\bibnamefont {van Troeye}}, \bibinfo
  {author} {\bibfnamefont {M.}~\bibnamefont {Lamparski}}, \bibinfo {author}
  {\bibfnamefont {V.}~\bibnamefont {Meunier}}, \bibinfo {author} {\bibfnamefont
  {V.-H.}\ \bibnamefont {Nguyen}}, \bibinfo {author} {\bibfnamefont
  {D.}~\bibnamefont {Paszko}}, \bibinfo {author} {\bibfnamefont {J.-C.}\
  \bibnamefont {Charlier}}, \bibinfo {author} {\bibfnamefont {L.~C.}\
  \bibnamefont {Campos}}, \bibinfo {author} {\bibfnamefont {L.~G.}\
  \bibnamefont {Can{\c{c}}ado}}, \bibinfo {author} {\bibfnamefont
  {G.}~\bibnamefont {Medeiros-Ribeiro}},\ and\ \bibinfo {author} {\bibfnamefont
  {A.}~\bibnamefont {Jorio}},\ }\href
  {https://doi.org/10.1038/s41586-021-03252-5} {\bibfield  {journal} {\bibinfo
  {journal} {Nature}\ }\textbf {\bibinfo {volume} {590}},\ \bibinfo {pages}
  {405} (\bibinfo {year} {2021})}\BibitemShut {NoStop}%
\bibitem [{\citenamefont {Urich}\ \emph {et~al.}(2011)\citenamefont {Urich},
  \citenamefont {Unterrainer},\ and\ \citenamefont {Mueller}}]{Urich2011}%
  \BibitemOpen
  \bibfield  {author} {\bibinfo {author} {\bibfnamefont {A.}~\bibnamefont
  {Urich}}, \bibinfo {author} {\bibfnamefont {K.}~\bibnamefont {Unterrainer}},\
  and\ \bibinfo {author} {\bibfnamefont {T.}~\bibnamefont {Mueller}},\ }\href
  {https://doi.org/10.1021/nl2011388} {\bibfield  {journal} {\bibinfo
  {journal} {NANO LETTERS}\ }\textbf {\bibinfo {volume} {11}},\ \bibinfo
  {pages} {2804} (\bibinfo {year} {2011})}\BibitemShut {NoStop}%
\bibitem [{\citenamefont {Sun}\ \emph {et~al.}(2012)\citenamefont {Sun},
  \citenamefont {Aivazian}, \citenamefont {Jones}, \citenamefont {Ross},
  \citenamefont {Yao}, \citenamefont {Cobden},\ and\ \citenamefont
  {Xu}}]{Sun2012}%
  \BibitemOpen
  \bibfield  {author} {\bibinfo {author} {\bibfnamefont {D.}~\bibnamefont
  {Sun}}, \bibinfo {author} {\bibfnamefont {G.}~\bibnamefont {Aivazian}},
  \bibinfo {author} {\bibfnamefont {A.~M.}\ \bibnamefont {Jones}}, \bibinfo
  {author} {\bibfnamefont {J.~S.}\ \bibnamefont {Ross}}, \bibinfo {author}
  {\bibfnamefont {W.}~\bibnamefont {Yao}}, \bibinfo {author} {\bibfnamefont
  {D.}~\bibnamefont {Cobden}},\ and\ \bibinfo {author} {\bibfnamefont
  {X.}~\bibnamefont {Xu}},\ }\href {https://doi.org/10.1038/NNANO.2011.243}
  {\bibfield  {journal} {\bibinfo  {journal} {NATURE NANOTECHNOLOGY}\ }\textbf
  {\bibinfo {volume} {7}},\ \bibinfo {pages} {114} (\bibinfo {year}
  {2012})}\BibitemShut {NoStop}%
\bibitem [{\citenamefont {Jadidi}\ \emph {et~al.}(2016)\citenamefont {Jadidi},
  \citenamefont {Suess}, \citenamefont {Tan}, \citenamefont {Cai},
  \citenamefont {Watanabe}, \citenamefont {Taniguchi}, \citenamefont {Sushkov},
  \citenamefont {Mittendorff}, \citenamefont {Hone}, \citenamefont {Drew},
  \citenamefont {Fuhrer},\ and\ \citenamefont {Murphy}}]{Jadidi2016}%
  \BibitemOpen
  \bibfield  {author} {\bibinfo {author} {\bibfnamefont {M.~M.}\ \bibnamefont
  {Jadidi}}, \bibinfo {author} {\bibfnamefont {R.~J.}\ \bibnamefont {Suess}},
  \bibinfo {author} {\bibfnamefont {C.}~\bibnamefont {Tan}}, \bibinfo {author}
  {\bibfnamefont {X.}~\bibnamefont {Cai}}, \bibinfo {author} {\bibfnamefont
  {K.}~\bibnamefont {Watanabe}}, \bibinfo {author} {\bibfnamefont
  {T.}~\bibnamefont {Taniguchi}}, \bibinfo {author} {\bibfnamefont {A.~B.}\
  \bibnamefont {Sushkov}}, \bibinfo {author} {\bibfnamefont {M.}~\bibnamefont
  {Mittendorff}}, \bibinfo {author} {\bibfnamefont {J.}~\bibnamefont {Hone}},
  \bibinfo {author} {\bibfnamefont {H.~D.}\ \bibnamefont {Drew}}, \bibinfo
  {author} {\bibfnamefont {M.~S.}\ \bibnamefont {Fuhrer}},\ and\ \bibinfo
  {author} {\bibfnamefont {T.~E.}\ \bibnamefont {Murphy}},\ }\bibfield
  {journal} {\bibinfo  {journal} {PHYSICAL REVIEW LETTERS}\ }\textbf {\bibinfo
  {volume} {117}},\ \href {https://doi.org/10.1103/PhysRevLett.117.257401}
  {10.1103/PhysRevLett.117.257401} (\bibinfo {year} {2016})\BibitemShut
  {NoStop}%
\bibitem [{\citenamefont {Aamir}\ \emph {et~al.}(2021)\citenamefont {Aamir},
  \citenamefont {Moore}, \citenamefont {Lu}, \citenamefont {Seifert},
  \citenamefont {Englund}, \citenamefont {Fong},\ and\ \citenamefont
  {Efetov}}]{Aamir2021}%
  \BibitemOpen
  \bibfield  {author} {\bibinfo {author} {\bibfnamefont {M.~A.}\ \bibnamefont
  {Aamir}}, \bibinfo {author} {\bibfnamefont {J.~N.}\ \bibnamefont {Moore}},
  \bibinfo {author} {\bibfnamefont {X.}~\bibnamefont {Lu}}, \bibinfo {author}
  {\bibfnamefont {P.}~\bibnamefont {Seifert}}, \bibinfo {author} {\bibfnamefont
  {D.}~\bibnamefont {Englund}}, \bibinfo {author} {\bibfnamefont {K.~C.}\
  \bibnamefont {Fong}},\ and\ \bibinfo {author} {\bibfnamefont {D.~K.}\
  \bibnamefont {Efetov}},\ }\href
  {https://doi.org/10.1021/acs.nanolett.1c01553} {\bibfield  {journal}
  {\bibinfo  {journal} {Nano Letters}\ }\textbf {\bibinfo {volume} {21}},\
  \bibinfo {pages} {5330} (\bibinfo {year} {2021})},\ \bibinfo {note} {pMID:
  34101476},\ \Eprint
  {https://arxiv.org/abs/https://doi.org/10.1021/acs.nanolett.1c01553}
  {https://doi.org/10.1021/acs.nanolett.1c01553} \BibitemShut {NoStop}%
\bibitem [{\citenamefont {Gabor}\ \emph {et~al.}(2011)\citenamefont {Gabor},
  \citenamefont {Song}, \citenamefont {Ma}, \citenamefont {Nair}, \citenamefont
  {Taychatanapat}, \citenamefont {Watanabe}, \citenamefont {Taniguchi},
  \citenamefont {Levitov},\ and\ \citenamefont {Jarillo-Herrero}}]{Gabor2011}%
  \BibitemOpen
  \bibfield  {author} {\bibinfo {author} {\bibfnamefont {N.~M.}\ \bibnamefont
  {Gabor}}, \bibinfo {author} {\bibfnamefont {J.~C.~W.}\ \bibnamefont {Song}},
  \bibinfo {author} {\bibfnamefont {Q.}~\bibnamefont {Ma}}, \bibinfo {author}
  {\bibfnamefont {N.~L.}\ \bibnamefont {Nair}}, \bibinfo {author}
  {\bibfnamefont {T.}~\bibnamefont {Taychatanapat}}, \bibinfo {author}
  {\bibfnamefont {K.}~\bibnamefont {Watanabe}}, \bibinfo {author}
  {\bibfnamefont {T.}~\bibnamefont {Taniguchi}}, \bibinfo {author}
  {\bibfnamefont {L.~S.}\ \bibnamefont {Levitov}},\ and\ \bibinfo {author}
  {\bibfnamefont {P.}~\bibnamefont {Jarillo-Herrero}},\ }\href
  {https://doi.org/10.1126/science.1211384} {\bibfield  {journal} {\bibinfo
  {journal} {Science}\ }\textbf {\bibinfo {volume} {334}},\ \bibinfo {pages}
  {648} (\bibinfo {year} {2011})},\ \Eprint
  {https://arxiv.org/abs/https://www.science.org/doi/pdf/10.1126/science.1211384}
  {https://www.science.org/doi/pdf/10.1126/science.1211384} \BibitemShut
  {NoStop}%
\bibitem [{\citenamefont {Ishizuka}\ \emph {et~al.}(2021)\citenamefont
  {Ishizuka}, \citenamefont {Fahimniya}, \citenamefont {Guinea},\ and\
  \citenamefont {Levitov}}]{Ishizuka2021}%
  \BibitemOpen
  \bibfield  {author} {\bibinfo {author} {\bibfnamefont {H.}~\bibnamefont
  {Ishizuka}}, \bibinfo {author} {\bibfnamefont {A.}~\bibnamefont {Fahimniya}},
  \bibinfo {author} {\bibfnamefont {F.}~\bibnamefont {Guinea}},\ and\ \bibinfo
  {author} {\bibfnamefont {L.}~\bibnamefont {Levitov}},\ }\href
  {https://doi.org/10.1021/acs.nanolett.1c00565} {\bibfield  {journal}
  {\bibinfo  {journal} {Nano Letters}\ }\textbf {\bibinfo {volume} {21}},\
  \bibinfo {pages} {7465} (\bibinfo {year} {2021})},\ \bibinfo {note} {pMID:
  34515488},\ \Eprint
  {https://arxiv.org/abs/https://doi.org/10.1021/acs.nanolett.1c00565}
  {https://doi.org/10.1021/acs.nanolett.1c00565} \BibitemShut {NoStop}%
\bibitem [{\citenamefont {Efetov}\ and\ \citenamefont
  {Kim}(2010)}]{Efetov2010}%
  \BibitemOpen
  \bibfield  {author} {\bibinfo {author} {\bibfnamefont {D.~K.}\ \bibnamefont
  {Efetov}}\ and\ \bibinfo {author} {\bibfnamefont {P.}~\bibnamefont {Kim}},\
  }\bibfield  {journal} {\bibinfo  {journal} {PHYSICAL REVIEW LETTERS}\
  }\textbf {\bibinfo {volume} {105}},\ \href
  {https://doi.org/10.1103/PhysRevLett.105.256805}
  {10.1103/PhysRevLett.105.256805} (\bibinfo {year} {2010})\BibitemShut
  {NoStop}%
\bibitem [{\citenamefont {Chen}\ \emph {et~al.}(2008)\citenamefont {Chen},
  \citenamefont {Jang}, \citenamefont {Xiao}, \citenamefont {Ishigami},\ and\
  \citenamefont {Fuhrer}}]{Chen2008}%
  \BibitemOpen
  \bibfield  {author} {\bibinfo {author} {\bibfnamefont {J.-H.}\ \bibnamefont
  {Chen}}, \bibinfo {author} {\bibfnamefont {C.}~\bibnamefont {Jang}}, \bibinfo
  {author} {\bibfnamefont {S.}~\bibnamefont {Xiao}}, \bibinfo {author}
  {\bibfnamefont {M.}~\bibnamefont {Ishigami}},\ and\ \bibinfo {author}
  {\bibfnamefont {M.~S.}\ \bibnamefont {Fuhrer}},\ }\href
  {https://doi.org/10.1038/nnano.2008.58} {\bibfield  {journal} {\bibinfo
  {journal} {NATURE NANOTECHNOLOGY}\ }\textbf {\bibinfo {volume} {3}},\
  \bibinfo {pages} {206} (\bibinfo {year} {2008})}\BibitemShut {NoStop}%
\bibitem [{\citenamefont {Dean}\ \emph {et~al.}(2010)\citenamefont {Dean},
  \citenamefont {Young}, \citenamefont {Meric}, \citenamefont {Lee},
  \citenamefont {Wang}, \citenamefont {Sorgenfrei}, \citenamefont {Watanabe},
  \citenamefont {Taniguchi}, \citenamefont {Kim}, \citenamefont {Shepard},\
  and\ \citenamefont {Hone}}]{Dean2010}%
  \BibitemOpen
  \bibfield  {author} {\bibinfo {author} {\bibfnamefont {C.~R.}\ \bibnamefont
  {Dean}}, \bibinfo {author} {\bibfnamefont {A.~F.}\ \bibnamefont {Young}},
  \bibinfo {author} {\bibfnamefont {I.}~\bibnamefont {Meric}}, \bibinfo
  {author} {\bibfnamefont {C.}~\bibnamefont {Lee}}, \bibinfo {author}
  {\bibfnamefont {L.}~\bibnamefont {Wang}}, \bibinfo {author} {\bibfnamefont
  {S.}~\bibnamefont {Sorgenfrei}}, \bibinfo {author} {\bibfnamefont
  {K.}~\bibnamefont {Watanabe}}, \bibinfo {author} {\bibfnamefont
  {T.}~\bibnamefont {Taniguchi}}, \bibinfo {author} {\bibfnamefont
  {P.}~\bibnamefont {Kim}}, \bibinfo {author} {\bibfnamefont {K.~L.}\
  \bibnamefont {Shepard}},\ and\ \bibinfo {author} {\bibfnamefont
  {J.}~\bibnamefont {Hone}},\ }\href {https://doi.org/10.1038/nnano.2010.172}
  {\bibfield  {journal} {\bibinfo  {journal} {NATURE NANOTECHNOLOGY}\ }\textbf
  {\bibinfo {volume} {5}},\ \bibinfo {pages} {722} (\bibinfo {year}
  {2010})}\BibitemShut {NoStop}%
\bibitem [{\citenamefont {Peltonen}\ \emph {et~al.}(2018)\citenamefont
  {Peltonen}, \citenamefont {Ojaj\"arvi},\ and\ \citenamefont
  {Heikkil\"a}}]{Peltonen2018}%
  \BibitemOpen
  \bibfield  {author} {\bibinfo {author} {\bibfnamefont {T.~J.}\ \bibnamefont
  {Peltonen}}, \bibinfo {author} {\bibfnamefont {R.}~\bibnamefont
  {Ojaj\"arvi}},\ and\ \bibinfo {author} {\bibfnamefont {T.~T.}\ \bibnamefont
  {Heikkil\"a}},\ }\href {https://doi.org/10.1103/PhysRevB.98.220504}
  {\bibfield  {journal} {\bibinfo  {journal} {Phys. Rev. B}\ }\textbf {\bibinfo
  {volume} {98}},\ \bibinfo {pages} {220504} (\bibinfo {year}
  {2018})}\BibitemShut {NoStop}%
\bibitem [{\citenamefont {Wallbank}\ \emph {et~al.}(2019)\citenamefont
  {Wallbank}, \citenamefont {Kumar}, \citenamefont {Holwill}, \citenamefont
  {Wang}, \citenamefont {Auton}, \citenamefont {Birkbeck}, \citenamefont
  {Mishchenko}, \citenamefont {Ponomarenko}, \citenamefont {Watanabe},
  \citenamefont {Taniguchi}, \citenamefont {Novoselov}, \citenamefont
  {Aleiner}, \citenamefont {Geim},\ and\ \citenamefont
  {Fal'ko}}]{Wallbank2019}%
  \BibitemOpen
  \bibfield  {author} {\bibinfo {author} {\bibfnamefont {J.~R.}\ \bibnamefont
  {Wallbank}}, \bibinfo {author} {\bibfnamefont {R.~K.}\ \bibnamefont {Kumar}},
  \bibinfo {author} {\bibfnamefont {M.}~\bibnamefont {Holwill}}, \bibinfo
  {author} {\bibfnamefont {Z.}~\bibnamefont {Wang}}, \bibinfo {author}
  {\bibfnamefont {G.~H.}\ \bibnamefont {Auton}}, \bibinfo {author}
  {\bibfnamefont {J.}~\bibnamefont {Birkbeck}}, \bibinfo {author}
  {\bibfnamefont {A.}~\bibnamefont {Mishchenko}}, \bibinfo {author}
  {\bibfnamefont {L.~A.}\ \bibnamefont {Ponomarenko}}, \bibinfo {author}
  {\bibfnamefont {K.}~\bibnamefont {Watanabe}}, \bibinfo {author}
  {\bibfnamefont {T.}~\bibnamefont {Taniguchi}}, \bibinfo {author}
  {\bibfnamefont {K.~S.}\ \bibnamefont {Novoselov}}, \bibinfo {author}
  {\bibfnamefont {I.~L.}\ \bibnamefont {Aleiner}}, \bibinfo {author}
  {\bibfnamefont {A.~K.}\ \bibnamefont {Geim}},\ and\ \bibinfo {author}
  {\bibfnamefont {V.~I.}\ \bibnamefont {Fal'ko}},\ }\href
  {https://doi.org/10.1038/s41567-018-0278-6} {\bibfield  {journal} {\bibinfo
  {journal} {NATURE PHYSICS}\ }\textbf {\bibinfo {volume} {15}},\ \bibinfo
  {pages} {32+} (\bibinfo {year} {2019})}\BibitemShut {NoStop}%
\bibitem [{\citenamefont {Jaoui}\ \emph {et~al.}(2022)\citenamefont {Jaoui},
  \citenamefont {Das}, \citenamefont {Di~Battista}, \citenamefont
  {Diez-Merida}, \citenamefont {Lu}, \citenamefont {Watanabe}, \citenamefont
  {Taniguchi}, \citenamefont {Ishizuka}, \citenamefont {Levitov},\ and\
  \citenamefont {Efetov}}]{Jaoui2022}%
  \BibitemOpen
  \bibfield  {author} {\bibinfo {author} {\bibfnamefont {A.}~\bibnamefont
  {Jaoui}}, \bibinfo {author} {\bibfnamefont {I.}~\bibnamefont {Das}}, \bibinfo
  {author} {\bibfnamefont {G.}~\bibnamefont {Di~Battista}}, \bibinfo {author}
  {\bibfnamefont {J.}~\bibnamefont {Diez-Merida}}, \bibinfo {author}
  {\bibfnamefont {X.}~\bibnamefont {Lu}}, \bibinfo {author} {\bibfnamefont
  {K.}~\bibnamefont {Watanabe}}, \bibinfo {author} {\bibfnamefont
  {T.}~\bibnamefont {Taniguchi}}, \bibinfo {author} {\bibfnamefont
  {H.}~\bibnamefont {Ishizuka}}, \bibinfo {author} {\bibfnamefont
  {L.}~\bibnamefont {Levitov}},\ and\ \bibinfo {author} {\bibfnamefont {D.~K.}\
  \bibnamefont {Efetov}},\ }\href {https://doi.org/10.1038/s41567-022-01556-5}
  {\bibfield  {journal} {\bibinfo  {journal} {NATURE PHYSICS}\ }\textbf
  {\bibinfo {volume} {18}},\ \bibinfo {pages} {633+} (\bibinfo {year}
  {2022})}\BibitemShut {NoStop}%
\bibitem [{\citenamefont {Ishizuka}\ and\ \citenamefont
  {Levitov}(2022)}]{Ishizuka2022}%
  \BibitemOpen
  \bibfield  {author} {\bibinfo {author} {\bibfnamefont {H.}~\bibnamefont
  {Ishizuka}}\ and\ \bibinfo {author} {\bibfnamefont {L.}~\bibnamefont
  {Levitov}},\ }\href {https://doi.org/10.1021/acs.nanolett.1c00565} {\bibfield
   {journal} {\bibinfo  {journal} {New J. Phys.}\ }\textbf {\bibinfo {volume}
  {24}},\ \bibinfo {pages} {052001} (\bibinfo {year} {2022})},\ \bibinfo {note}
  {pMID: 34515488},\ \Eprint
  {https://arxiv.org/abs/https://doi.org/10.1088/1367-2630/ac688c}
  {https://doi.org/10.1088/1367-2630/ac688c} \BibitemShut {NoStop}%
\bibitem [{\citenamefont {Di~Battista}\ \emph {et~al.}(2022)\citenamefont
  {Di~Battista}, \citenamefont {Seifert}, \citenamefont {Watanabe},
  \citenamefont {Taniguchi}, \citenamefont {Fong}, \citenamefont {Principi},\
  and\ \citenamefont {Efetov}}]{DiBattista2022}%
  \BibitemOpen
  \bibfield  {author} {\bibinfo {author} {\bibfnamefont {G.}~\bibnamefont
  {Di~Battista}}, \bibinfo {author} {\bibfnamefont {P.}~\bibnamefont
  {Seifert}}, \bibinfo {author} {\bibfnamefont {K.}~\bibnamefont {Watanabe}},
  \bibinfo {author} {\bibfnamefont {T.}~\bibnamefont {Taniguchi}}, \bibinfo
  {author} {\bibfnamefont {K.~C.}\ \bibnamefont {Fong}}, \bibinfo {author}
  {\bibfnamefont {A.}~\bibnamefont {Principi}},\ and\ \bibinfo {author}
  {\bibfnamefont {D.~K.}\ \bibnamefont {Efetov}},\ }\href
  {https://doi.org/10.1021/acs.nanolett.1c04512} {\bibfield  {journal}
  {\bibinfo  {journal} {Nano Letters}\ }\textbf {\bibinfo {volume} {22}},\
  \bibinfo {pages} {6465} (\bibinfo {year} {2022})},\ \bibinfo {note} {pMID:
  35917225},\ \Eprint
  {https://arxiv.org/abs/https://doi.org/10.1021/acs.nanolett.1c04512}
  {https://doi.org/10.1021/acs.nanolett.1c04512} \BibitemShut {NoStop}%
\bibitem [{\citenamefont {Deng}\ \emph {et~al.}(2020)\citenamefont {Deng},
  \citenamefont {Ma}, \citenamefont {Wang}, \citenamefont {Yuan}, \citenamefont
  {Watanabe}, \citenamefont {Taniguchi}, \citenamefont {Zhang},\ and\
  \citenamefont {Xia}}]{Deng2020}%
  \BibitemOpen
  \bibfield  {author} {\bibinfo {author} {\bibfnamefont {B.}~\bibnamefont
  {Deng}}, \bibinfo {author} {\bibfnamefont {C.}~\bibnamefont {Ma}}, \bibinfo
  {author} {\bibfnamefont {Q.}~\bibnamefont {Wang}}, \bibinfo {author}
  {\bibfnamefont {S.}~\bibnamefont {Yuan}}, \bibinfo {author} {\bibfnamefont
  {K.}~\bibnamefont {Watanabe}}, \bibinfo {author} {\bibfnamefont
  {T.}~\bibnamefont {Taniguchi}}, \bibinfo {author} {\bibfnamefont
  {F.}~\bibnamefont {Zhang}},\ and\ \bibinfo {author} {\bibfnamefont
  {F.}~\bibnamefont {Xia}},\ }\href {https://doi.org/10.1038/s41566-020-0644-7}
  {\bibfield  {journal} {\bibinfo  {journal} {NATURE PHOTONICS}\ }\textbf
  {\bibinfo {volume} {14}},\ \bibinfo {pages} {549+} (\bibinfo {year}
  {2020})}\BibitemShut {NoStop}%
\end{thebibliography}%

%\end{document}

%\begin{appendices}
%\pagenumbering{gobble}
\newpage
\clearpage

\ExtendedDataFigures

\section{Extended Data Figures}\label{secEDF}

\begin{figure*}[ht]%
	\centering
	\includegraphics[width=\textwidth,height=\textheight,keepaspectratio]{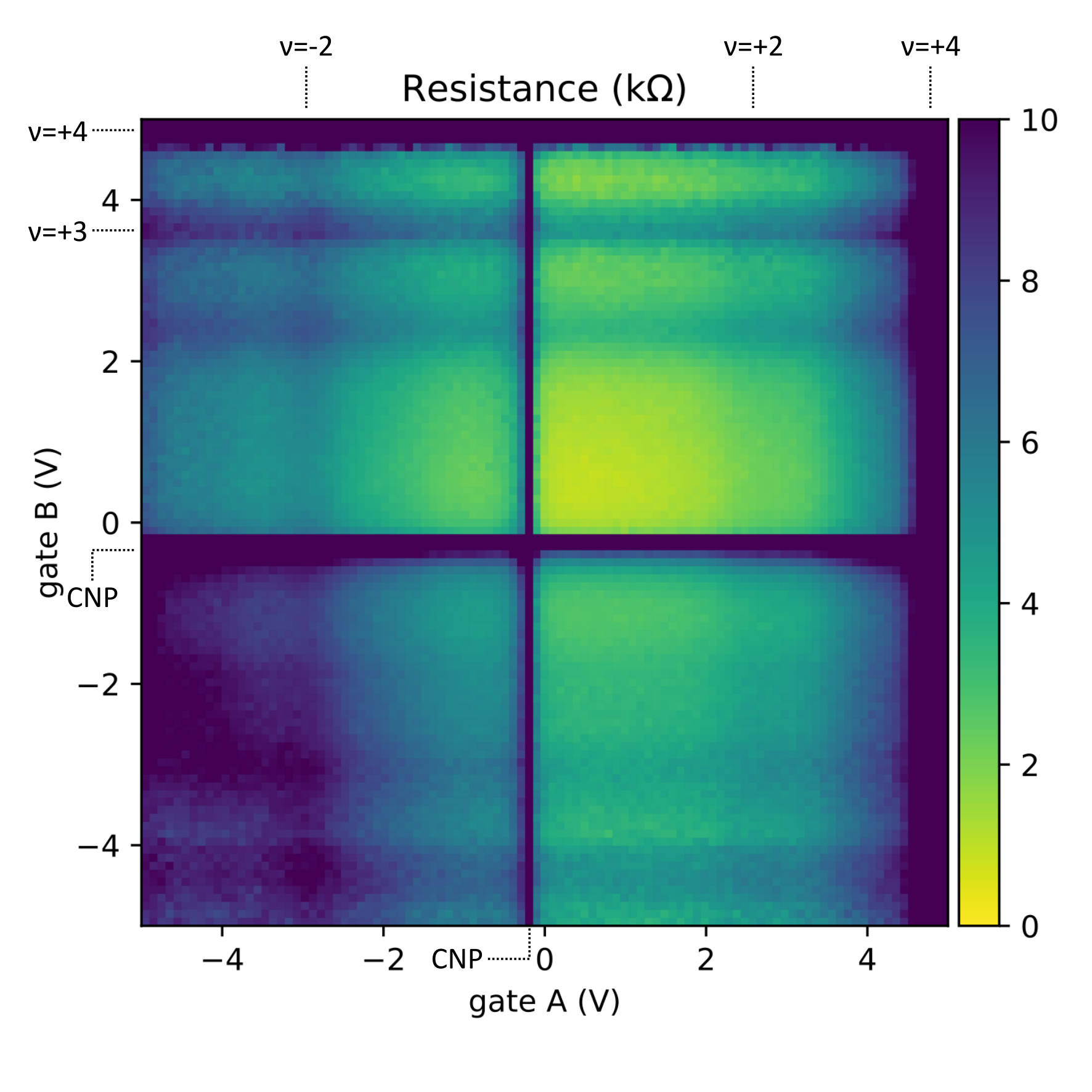}
	\caption{Dual gate map of the four-probe resistance of MATBG ($\theta=1.24^\circ$) at T=3.6 K. The maxima in resistance correspond to the charge neutrality points (CNPs) and integer filling factors ($\nu=\pm2,\pm3,\pm4$).}\label{edf1}
\end{figure*}

\begin{figure*}[ht]%
	\centering
	\includegraphics[width=0.95\textwidth,height=0.95\textheight,keepaspectratio]{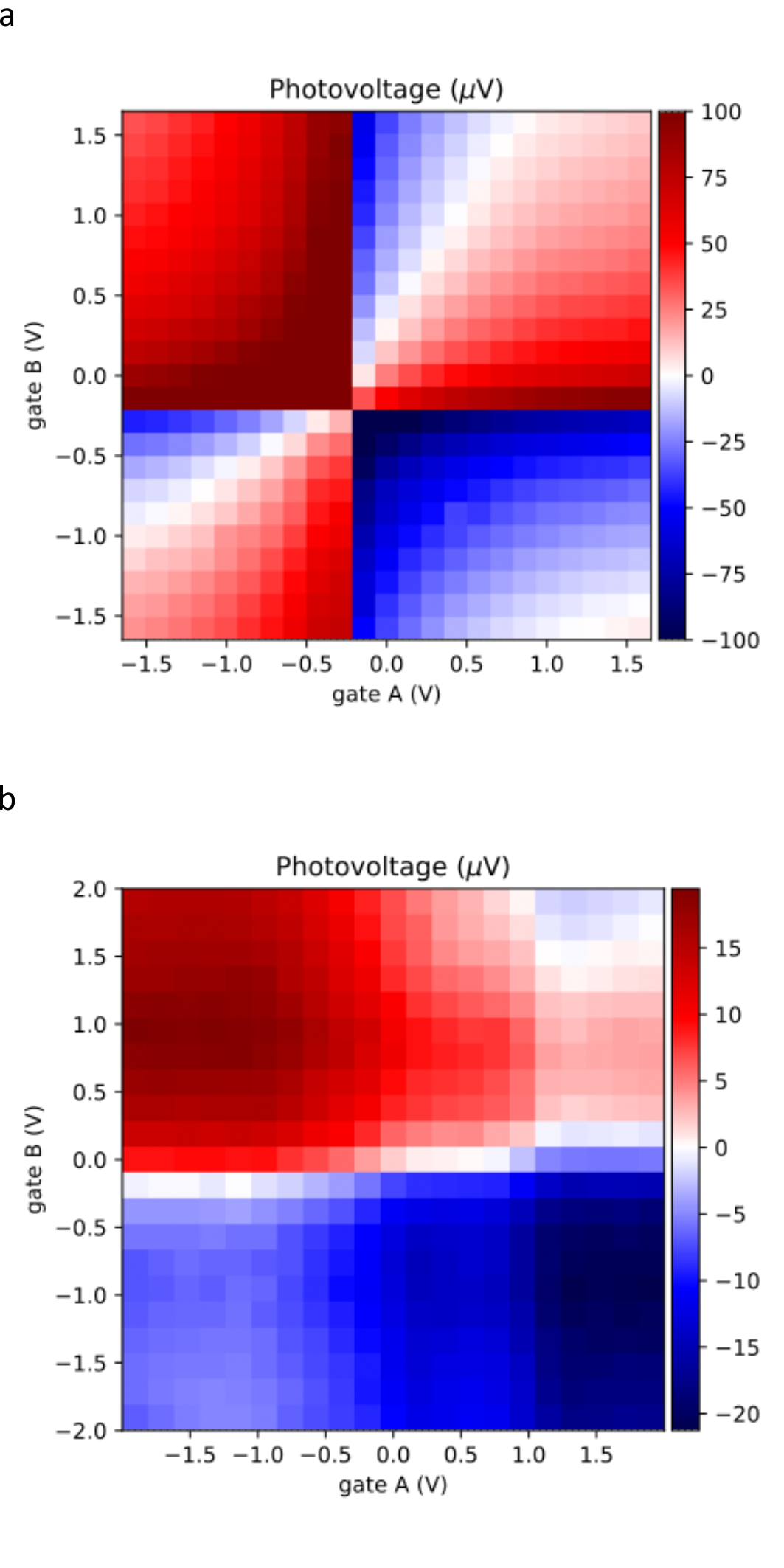}
	\caption{Dual gate photovoltage maps for; \textbf{a} MATBG ($\theta=1.24^\circ$, $T=10$ K) and \textbf{b} BLG ($\theta=0^\circ$, $T=100$ K).}\label{edf2}
\end{figure*}

\begin{figure*}[ht]%
	\centering
	\includegraphics[width=\textwidth,height=\textheight,keepaspectratio]{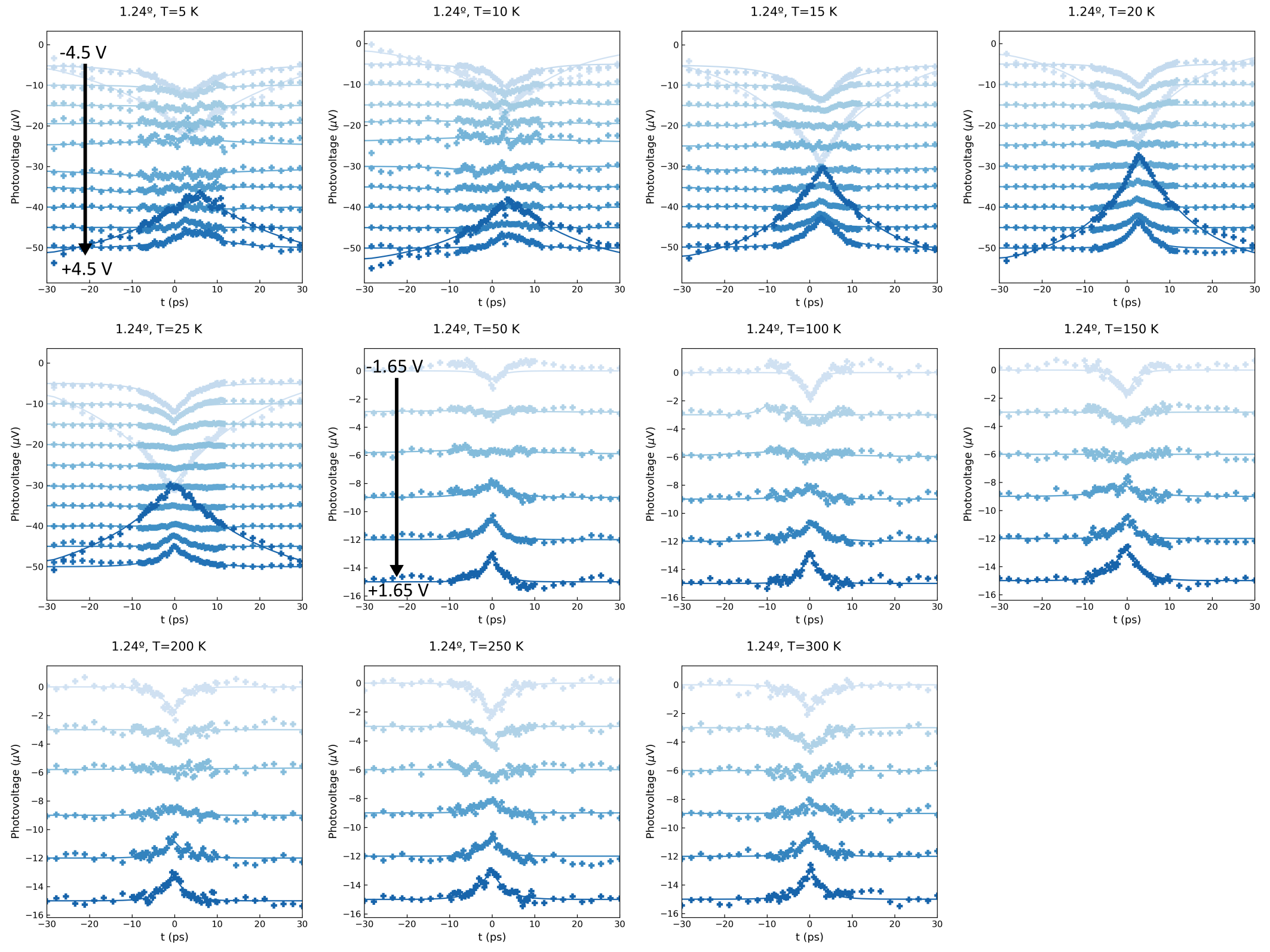}
	\caption{TrPV dips for the MATBG ($\theta=1.24^\circ$) device as a function of DU vector (indicated by arrow) and temperature (see plot title). Each time trace has been offset for clarity.}\label{edf3}
\end{figure*}

\begin{figure*}[ht]%
	\centering
	\includegraphics[width=\textwidth,height=\textheight,keepaspectratio]{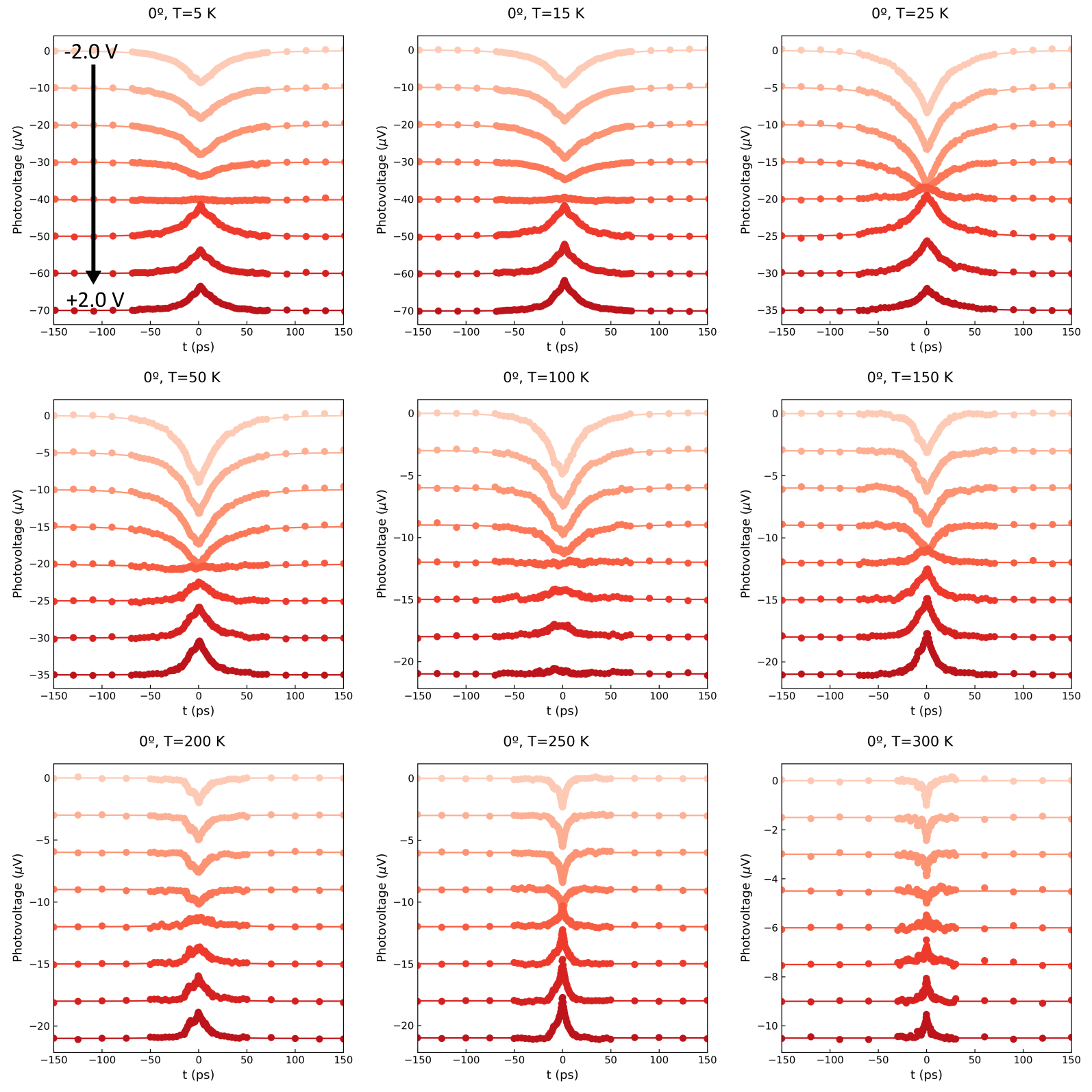}
	\caption{TrPV dips for the BLG ($\theta=0^\circ$) device as a function of DU vector (indicated by arrow) and temperature (see plot title). Each time trace has been offset for clarity. The slower cooling at low temperatures produces a broader dip.}\label{edf4}
\end{figure*}

\begin{figure*}[ht]%
	\centering
	\includegraphics[width=\textwidth,height=\textheight,keepaspectratio]{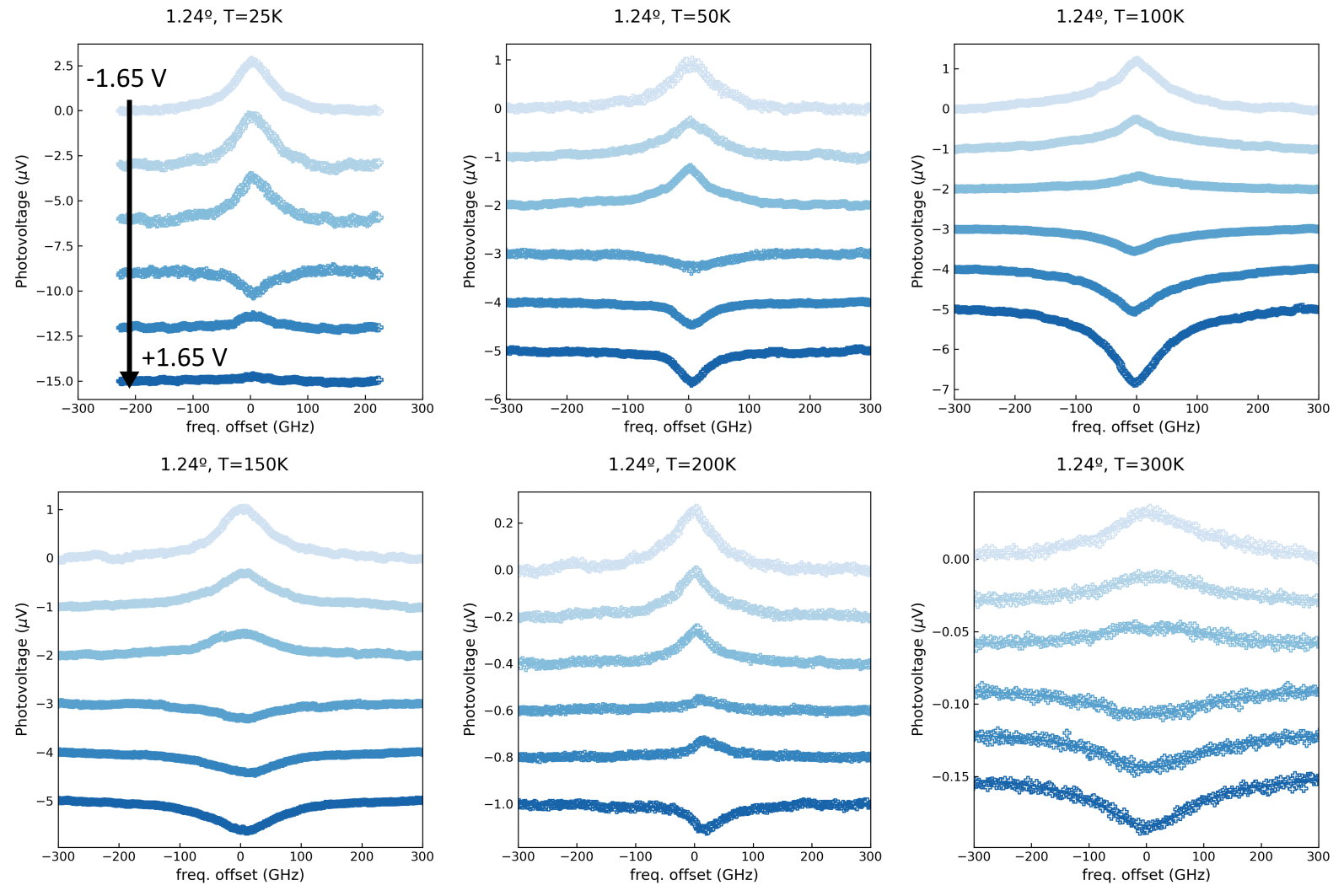}
	\caption{CW-PM peaks for the MATBG ($\theta=1.24^\circ$) device as a function of DU vector (indicated by arrow) and temperature (see plot title). Each frequency sweep has been offset for clarity.}\label{edf5}
\end{figure*}

\begin{figure*}[ht]%
	\centering
	\includegraphics[width=\textwidth,height=\textheight,keepaspectratio]{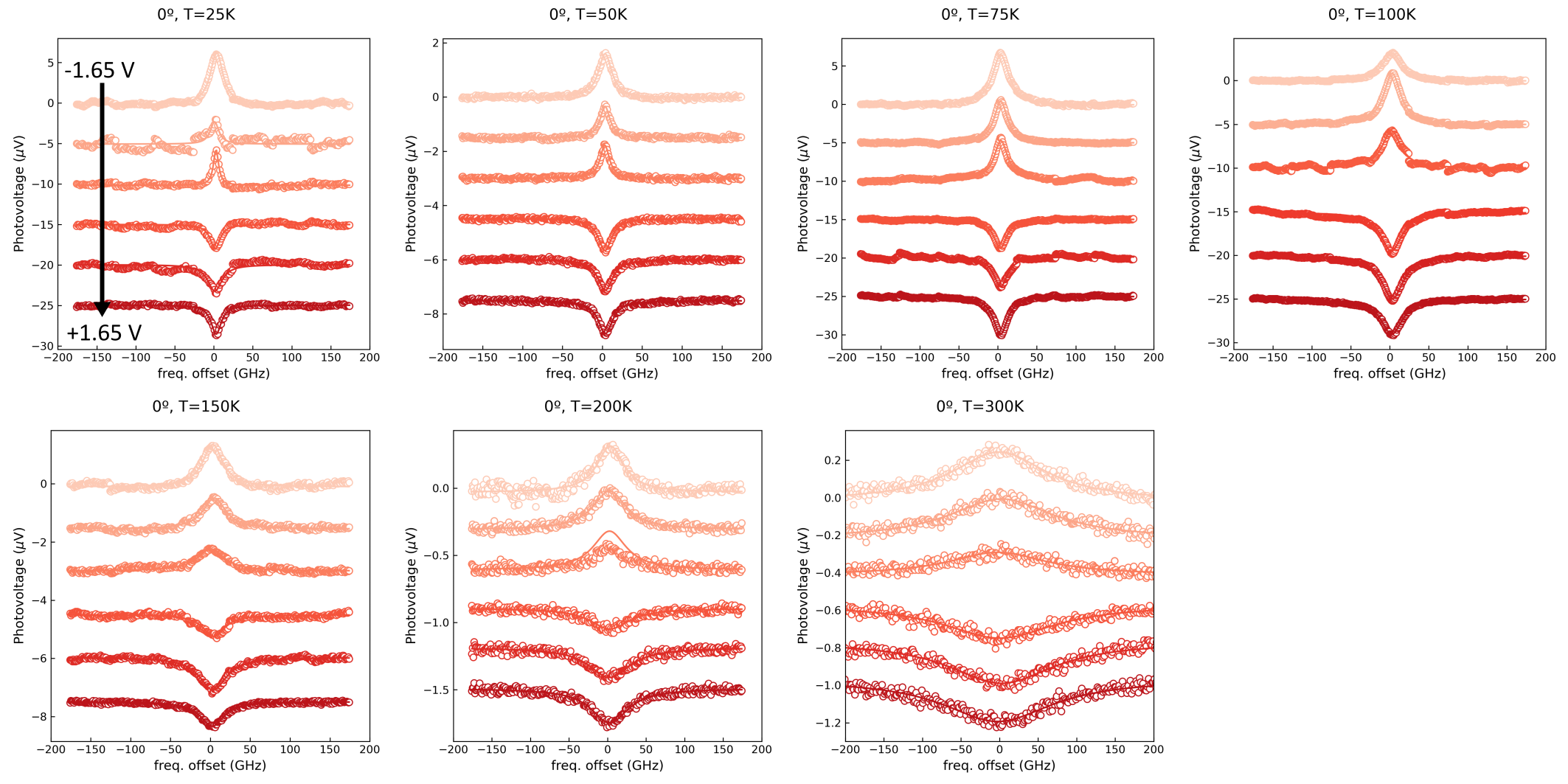}
	\caption{CW-PM peaks for the BLG ($\theta=0^\circ$) device as a function of DU vector (indicated by arrow) and temperature (see plot title). Each frequency sweep has been offset for clarity. The slower cooling at low temperatures produces a narrower peak.}\label{edf6}
\end{figure*}

\begin{figure*}[ht]%
	\centering
	\includegraphics[width=\textwidth,height=\textheight,keepaspectratio]{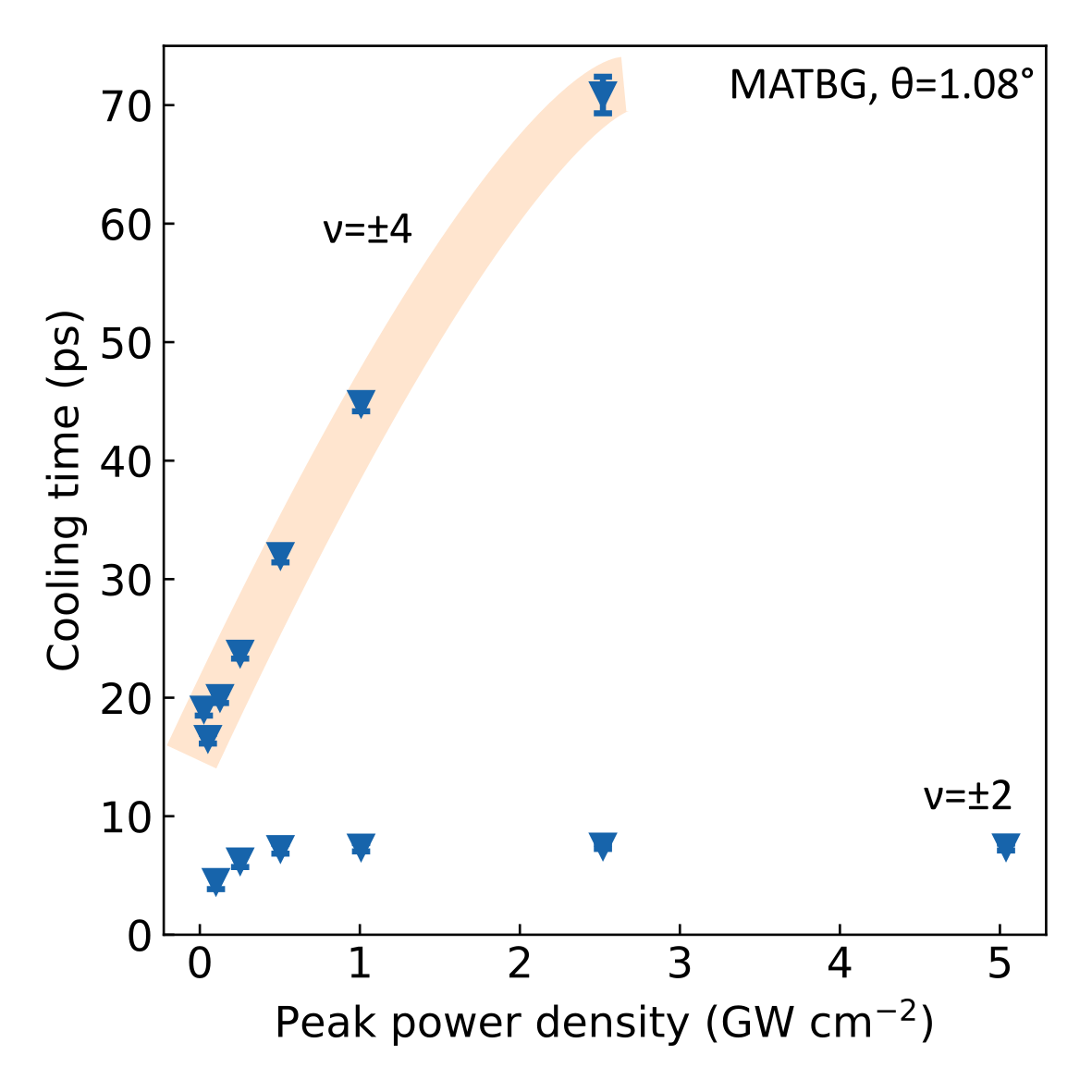}
	\caption{Power dependence of cooling time for a second MATBG device ($\theta=1.08^\circ$). The electron relaxation bottleneck at full filling ($\nu=\pm4$) leads to slower cooling time for higher laser powers. The orange line is a guide to eye.}\label{edf7}
\end{figure*}

\newpage
\clearpage

\SupplementaryMaterials

\section{Supplementary Information}\label{secSI}%
\subsection*{Supplementary Note 1}\label{secSI1}
In Supp. Fig. \ref{SI:figS3}, we investigate the influence of twist angle disorder on the electrical transport at $T=35$ mK. At the junction contact, the twist angle is $\theta=1.24^\circ$ and we observe sharp resistance peaks at $\nu=\pm2$ arising from correlated insulating states. The contacts at the top of the junction display a shoulder around $\nu=-2$ that indicates a mixing of two angles ($\theta=1.24-1.28^\circ$). For the contacts at the bottom of the junction the angle is $\theta=1.24^\circ$. From this we conclude that there is minimal twist angle disorder in the proximity of the pn-junction. 

\begin{figure}[ht]%
	\centering
	\includegraphics[width=0.7\textwidth]{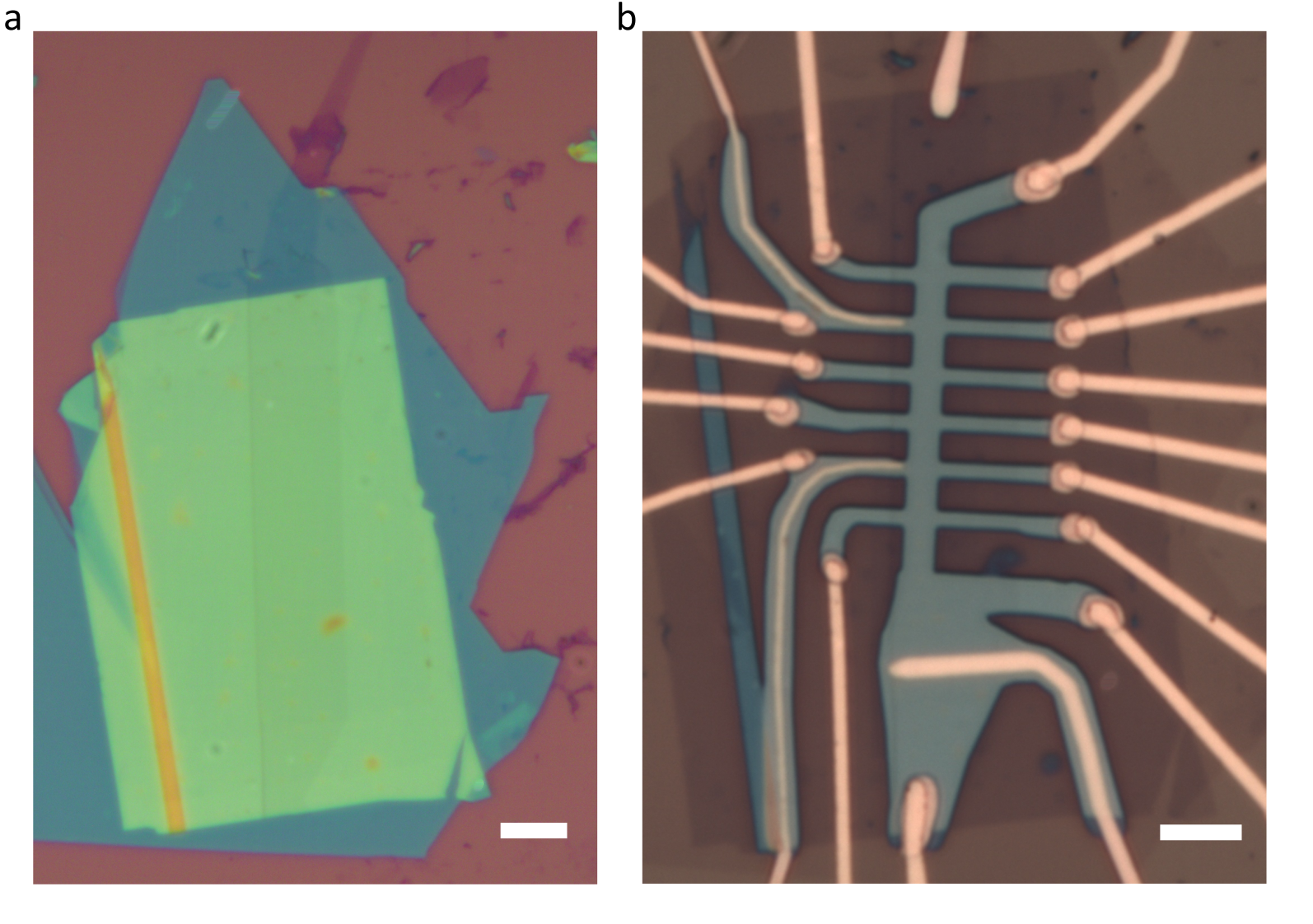}
	\caption{Optical images of the device before and after nanofabrication. \textbf{a}, Heterostructure stack dropped on a Si/SiO$_2$ substrate. \textbf{b}, Finalised device after etching Hall bar and metallisation. Both scale bars are $5\mu$m.}\label{SI:figS1}
\end{figure}

\begin{figure*}[ht]%
	\centering
	\includegraphics[width=0.7\textwidth]{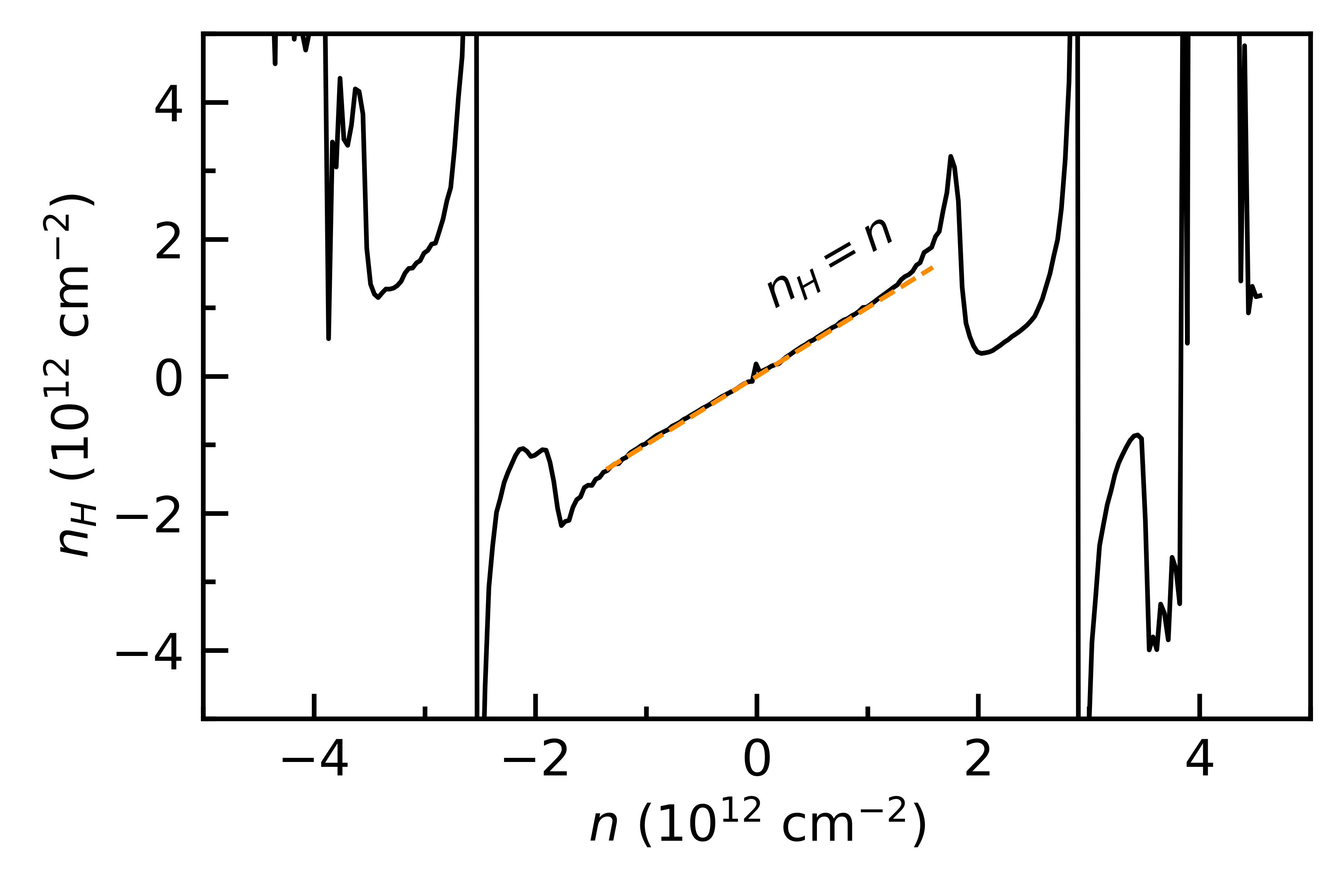}
	\caption{Low field Hall effect at 1.8 K. Hall carrier density $n_H$ vs $n$. In the region close to charge neutrality $n_H = n$, which allows us to calibrate the relationship between $V_g$ and $n$ to extract the twist angle.}\label{SI:figS2}
\end{figure*}
\begin{figure*}[ht]%
	\centering
	\includegraphics[width=0.7\textwidth]{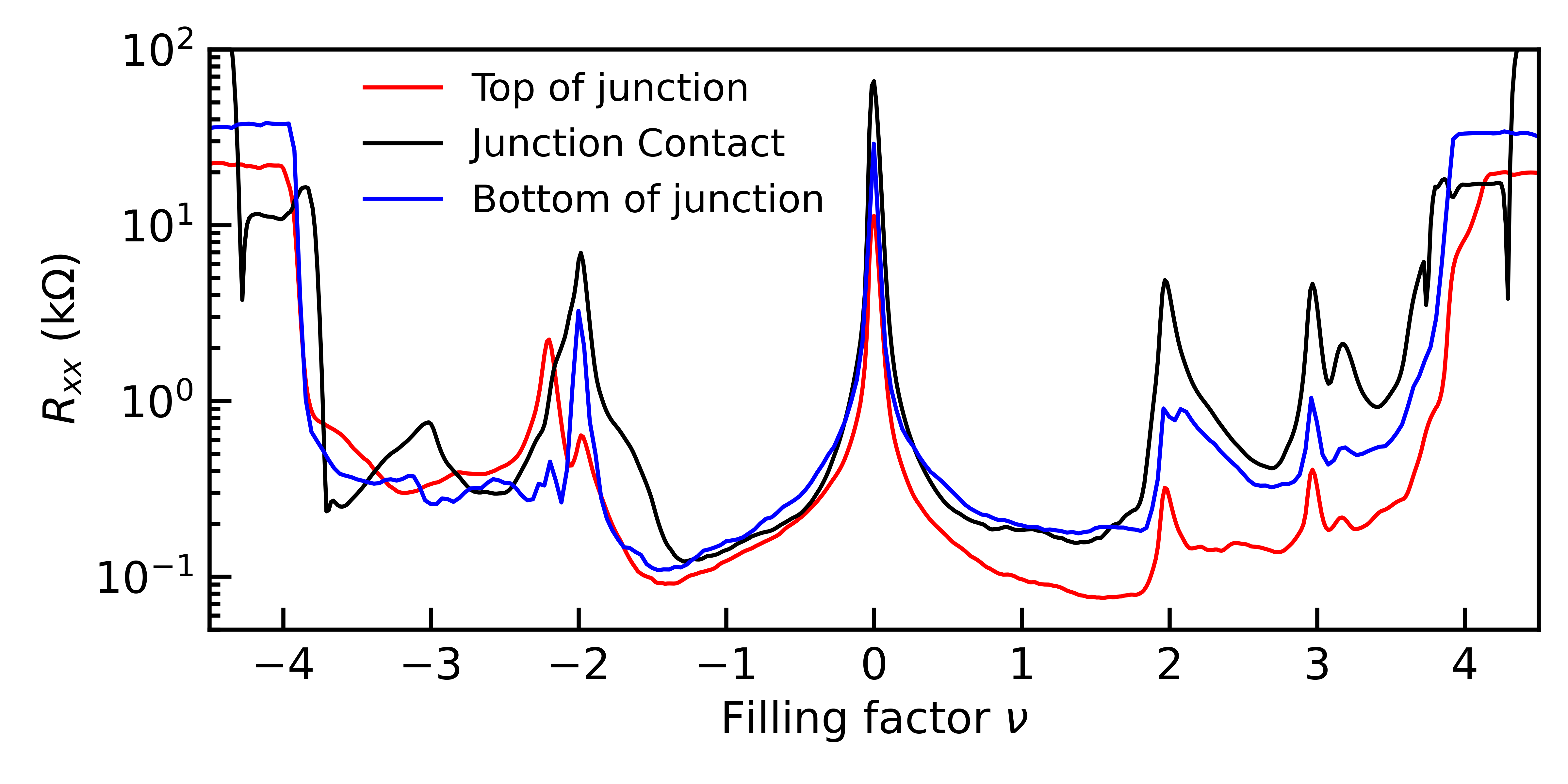}
	\caption{Longitudinal resistance ($R_{xx}$) vs. filling factor ($\nu$) for contacts around the junction.}\label{SI:figS3}
\end{figure*}

\end{document}